\documentclass[twocolumn]{raa}            % referee version: for submission
\usepackage{graphicx, times}             %for PS/EPS graphics inclusion, new
\usepackage{natbib}
\usepackage{amssymb,amsmath}
\usepackage{CJKutf8}
\bibpunct{(}{)}{;}{a}{}{,}
\usepackage{diagbox}
\usepackage{appendix}
\usepackage{threeparttable}
\usepackage{xcolor}
\usepackage{amsmath}
\usepackage{color}
\usepackage{subcaption}
\usepackage{authblk}
\usepackage{enumitem} % 引入 enumitem 宏包
\usepackage{wrapfig}  % 新增：实现图片悬浮在文字上的宏包
\usepackage{float}    % 辅助控制图片浮动

\usepackage{url}

\usepackage[a4paper=true,driverfallback=dvipdfm,pagebackref=true]{hyperref}
\hypersetup{colorlinks = true, linkcolor = green, anchorcolor = red, citecolor = blue, filecolor = red, pagecolor = red, urlcolor = red}
\usepackage{threeparttable}
\usepackage{orcidlink}

\mathchardef\mhyphen="2D

\begin{document}
% \begin{CJK}{UTF8}{gbsn}

   \title{The Average Age Map of the Galactic Bulge}

   \volnopage{Vol.0 (20xx) No.0, 000--000}      %%preserved for Editor. Don't remove it!
   \setcounter{page}{1}          %%starting page, preserved for Editor. Don't remove it!

\author{Jialu Nie~\orcidlink{0000-0002-1418-9081}\inst{1,2}, Mart\'\i n L\'opez-Corredoira~\orcidlink{0000-0001-6128-6274}\inst{3,4,5}, Chao Liu~\orcidlink{0000-0002-1802-6917}\inst{1,2,6,7}\thanks{Corresponding author: liuchao@nao.cas.cn}, Hai-feng Wang~\orcidlink{0000-0001-8459-1036}\inst{8,9,10} and Iulia Simion~\orcidlink{0000-0001-8889-0762}\inst{11}}
% \email{liuchao@nao.cas.cn}

\institute{National Astronomical Observatories, Chinese Academy of Sciences, Beijing 100101, People's Republic of China \\
\and
School of Astronomy and Space Science, University of Chinese Academy of Sciences, Beijing 100049, People's Republic of China \\
\and
Instituto de Astrofísica de Canarias, E-38205 La Laguna, Tenerife, Spain\\
\and
Departamento de Astrofísica, Universidad de La Laguna, E-38206 La Laguna, Tenerife, Spain\\
\and
PIFI-Visiting Scientist 2023 of Chinese Academy of Sciences at Purple Mountain Observatory, Nanjing 210023, and National Astronomical Observatories, Beijing 100012, People's Republic of China\\
\and
Institute for Frontiers in Astronomy and Astrophysics, Beijing Normal University, Beijing 100875, People's Republic of China\\
\and 
Zhejiang Lab, Hangzhou 311121, People's Republic of China\\
\and
Local Universe and Time-Domain Astronomy Laboratory, China West Normal University, Nanchong 637002, People's Republic of China\\
\and
Luoxiahong Institute of Astronomy, China West Normal University, Nanchong 637002, People's Republic of China\\
\and
Department of Astronomy, China West Normal University, Nanchong 637002, People's Republic of China
\and
Shanghai Key Laboratory for Astrophysics, Shanghai Normal University, 100 Guilin Road, Shanghai 200234, People's Republic of China
}

% \correspondingauthor{liuchao@nao.cas.cn}

\abstract{
The Galactic Bulge, as the center of the Galaxy, is the closest laboratory for studying galaxy formation and evolution. However, its study faces significant challenges due to heavy dust extinction. This paper is devoted to deriving the average age of the Galactic Bulge and investigating its spatial distribution. We utilize a high-precision PSF-fitting photometric catalogue in the $J$ and $K_{\mathrm{s}}$ bands observed by VISTA to study the average stellar ages within the Bulge. Red giant stars are employed as tracers, with their average distances determined using red clump stars as references. The average ages are fitted with stellar models. 
Our analysis reveals a systematic age gradient across the Galactic Bulge ($2^{\circ} < |b| < 8^{\circ}$). The mean stellar age increases significantly with galactic latitude, shifting from a younger population ($\sim 4.69^{+0.97}_{-0.81}$ Gyr) prevalent near the plane to a predominantly older population ($\sim 10.48^{+0.93}_{-0.85}$ Gyr) at higher latitudes.
We hypothesize that the young stellar population at low latitudes is predominantly composed of a pseudo-bulge formed via disk/bar processes (incorporating contributions from recent star-forming activity in the Galactic center), whereas the older stellar population is associated with spheroidal bulges generated through early-stage collapse or accretion of debris from merged dwarf galaxies.  
    }
   \authorrunning{Nie et al.}            %author_head in even pages
   \titlerunning{The Average Age Map of the Galactic Bulge}  % title_head in odd pages

   \maketitle

\section{Introduction} 
\label{sec:intro}
The interior of galaxies preserves remnants of their earliest activities, including ancient stellar populations formed during the primordial stages. Therefore, these regions are essential for unravelling the history of the Galaxy's formation.

The morphology of the Galactic component has been well studied over the past few decades, following the advent of near-infrared surveys. \citet{Wei94} first revealed the boxy nature of the Galactic Bulge using COBE satellite images. Subsequently, several researchers analysed the density of red clump (RC) stars \citep[e.g.,][]{Nat10,cao_new_2013} and suggested that the Bulge exhibits an X-shaped structure formed by eccentric orbits.
\citet{portail_made--measure_2015} characterised this X-shape using orbital parameters and determined that stars forming the X-structure account for 24\% of the Bulge's mass. \citet{ness_x-shaped_2016} later confirmed the X-shaped morphology of the Galactic Bulge through WISE images. This observation provides critical insights into the formation of the Bulge, particularly in models involving buckling instabilities \citep{Li15}.
However, the X-structure does not show up in other populations but only in the metal-rich RC stars \citep{Lop17}, which raises suspicions about potential problems in the density measurements using RC stars. In fact, evidence suggests that main RC stars are contaminated by a secondary red-clump in the Bulge \citep{lopez-corredoira_distribution_2019}, leading to double peaks in stellar counts but not in density distributions along the lines of sight. Furthermore, images provided by ~\cite{ness_x-shaped_2016} indicate that the X-shape might be an artifact resulting from the subtraction of specific disk models or assuming the Bulge as an ellipsoid rather than a boxy structure \citep{Lop17,Han18}. \citet{han_structure_2025} also revealed that the Boxy-Peanut (B/P) bulge may be more dominant than the X-shaped bulge. 

Studying the stellar metallicity in the Galactic Bulge can further constrain its structure and evolution. Stellar spectroscopic studies have revealed that the metallicity distribution in this region can be divided into metal-poor (MP) stars and metal-rich (MR) stars. \cite{zoccali_giraffe_2017} obtained the metallicity distributions for these two populations using spectroscopic data of RC stars from multiple regions within the Galactic Bulge and discussed their spatial distribution and evolutionary implications. They found that MP stars dominate in the outer regions, exhibiting a spherical distribution; the MR population is concentrated towards the plane, demonstrating a boxy distribution. \cite{rojas-arriagada_how_2020} fitted the metallicities of 13,000 stars from APOGEE using three Gaussian distributions, which represent the metal-poor, metal-intermediate, and metal-rich populations, respectively. They confirmed that the metal-rich stars are associated with the long-term evolution of the square/peanut-shaped bars in the disk; the metal-intermediate stars may represent products formed in situ at high redshift; and the metal-poor stars are likely linked to the metal-poor tail of the early thick disk. Studies of metallicity distribution illustrate that stellar populations in the Bulge region have a complex composition, with clear gradients depending on the location. For example, a global 2D map of metallicity has been provided by ~\citet{Joh22}.

Age-related studies in the Bulge region are more directly linked to understanding the history of the Bulge formation in terms of an old spheroidal population and a younger pseudobulge associated with the buckling instability of the disc:
% \begin{itemize}
\begin{itemize}[label=\textbullet]
\item On one hand, 
\cite{ortolani_near-coeval_1995} studied two metal-rich globular clusters from the Hubble Space Telescope (HST) observations in the Bulge, NGC6528 and NGC6553, and found that they are older than 10 Gyr. In Baade's window ($l=1.1^\circ,b=-4.8^\circ$), 
the luminosity function drops down for very bright stars: $M_V<4.0$ mag, $M_I<3.5$ mag \citep{Hol98},  
as expected for an old population. \cite{spite_age_1999} concluded, through testing of Baade's window and Sagittarius-I ($l=1.3^\circ,b=-2.7^\circ$), that the Bulge should be dominated by the old population. \cite{zoccali_age_2003} used the photometric data from the 2MASS survey and found that the age of the Bulge is as large as that of Galactic globular clusters, which is $\gtrsim$ 10 Gyr. Combined with the UV-to-IR stellar photometry of four low-extinction windows with HST, \cite{brown_wfc3_2010} found that these fields are all dominated by old populations ($\sim$ 10 Gyr). \cite{clarkson_first_2011} first detected Blue Straggler Stars (BSS) in the SWEEPS (Sagittarius Window Eclipsing Extrasolar Planet Search) field ($l=2.65^\circ,b=-1.25^\circ$), but they found that the genuinely young population ($<$ 5 Gyr) accounted for a maximum of 3.4\% according to the estimation of the number of BSS in the Bulge. \cite{valenti_stellar_2013} calculated the age of the two fields in the Southern region based on the near-IR images of the VLT/HAWK-I and also concluded that their ages are $>$10 Gyr. \cite{surot_mapping_2019} performed age determinations of stellar populations in the field b249 ($l=-0.45^\circ,b=-6.39^\circ$) of the VISTA Variable in the Via Lactea (VVV) survey. They found that most of the stars are older than 7.5 Gyr, and the age range from 7.5 Gyr to 11 Gyr.

\item On the other hand, Mira stars with the longest periods (age$\lesssim $5 Gyr) show clear evidence of a barred structure of the Bulge \citep{Whi91, Cat16}. The presence of infrared carbon stars indicates the presence of a significant population younger than 6 Gyr \citep{Col02}. \cite{bensby_chemical_2013} suggested a large span of stellar ages (2 Gyr-12 Gyr) in the Galactic Bulge based on high-resolution spectra obtained during gravitational microlensing events. Furthermore, \cite{Ben17} found that $>$18\% of the stars in the Bulge are younger than 5 Gyr (spectroscopic ages). \cite{haywood_hiding_2016} and \cite{bernard_star_2018} used the systematic stellar evolution template instead of isochrones to fit the observed data of HST (also HST windows, which are SWEEPS, Stanek, Baade and OGLE29) and found similar results to those of \cite{bensby_chemical_2013}. \cite{schultheis_baades_2017} and \cite{hasselquist_exploring_2020} used the spectroscopic data from the Apache Point Observatory Galactic Evolution Experiment (APOGEE) to determine the stellar age and metallicity. They found the presence of a non-negligible number of young stars (age $\sim$2 Gyr-5 Gyr) in the Bulge. Furthermore, the radial velocity distribution derived from the Gaia DR3 data may indicate ages lower than 5 Gyr \citep{melnik_humps_2024}.
\end{itemize}

There is a connection between age and metallicity: Bulge dwarf stars have consistently old ages at sub-solar metallicities, while at super-solar metallicities they cover a wide range of ages \citep{bensby_chemical_2010}. Given the observed metallicity gradients, one may expect corresponding gradients in age. However, many studies on the age of the Galactic Bulge have been limited to small regions with sparse samples due to the challenging environment of the Galactic Bulge. While some trends suggest a younger population near the plane \citep{Nes14, Ben17}, there is still no global map of the Bulge's average age.
The studies mentioned above, based on photometric data, have yielded ages older than 10 Gyr; however, spectroscopic studies have shown discrepancies. We suggest that these earlier studies may suffer from selection biases \citep{queiroz_bulge_2020, queiroz_milky_2021} due to their reliance on the low-extinction windows in HST observations or small fields within the Bulge. 
To address this, we present a map of the Galactic Bulge's average age, leveraging photometric data from as many sky regions in the VVV survey as possible to achieve a statistical result rather than focusing on individual stars. We aim to derive the average age of almost the entire Bulge and analyze its distribution. While our work is not without observational and modeling limitations, it still captures some key trends, as shown in Figure \ref{fig:4.2-3}.

In this paper, we first describe the dataset that we used in Section~\ref{sec:Data}. Then, we describe the methodology to determine the average age in Section~\ref{sec:Mth} and analyse the average age for individual fields and the whole focused region in Section~\ref{sec:Re}.  
In Section~\ref{sec:Dis}, we evaluate the impacts of distance uncertainties, foreground contamination, and sample selection biases. We also explore the potential influence of stellar model assumptions and outline the limitations of this work.
Finally, we provide a summary in Section~\ref{sec:Sum}.

\section{Dataset} \label{sec:Data}
\begin{figure*}[ht!]
\centering
\includegraphics[scale=0.5]{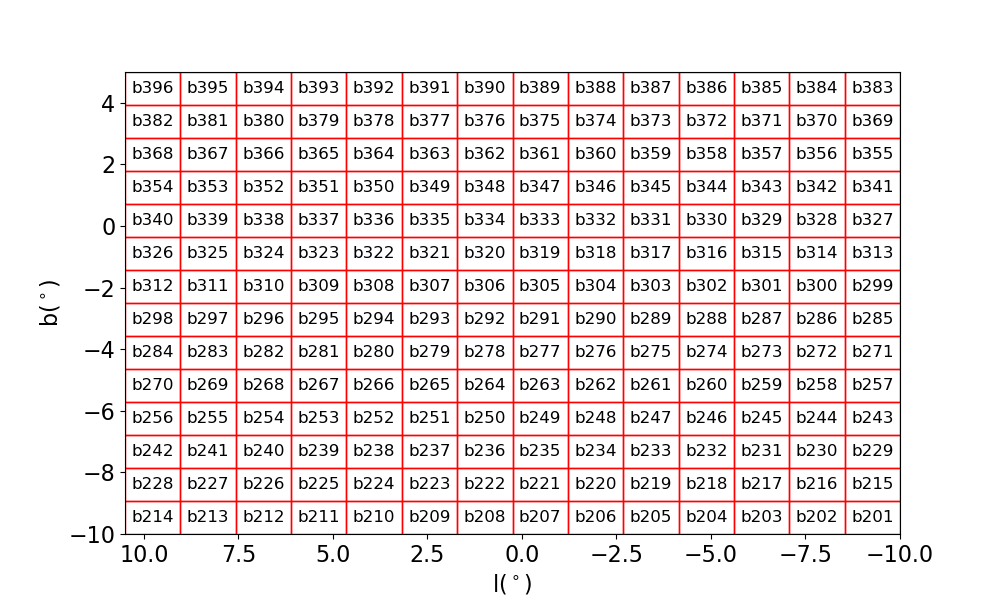}
\caption{Schematic diagram of the Bulge region in the VVV survey, with numbers indicating the tile IDs.
}
\label{fig:2.1}
\end{figure*}

The VVV survey \citep{minniti_vista_2010} is the public ESO near-IR variability survey which covers the Milky Way Bulge ($-10^{\circ} \leq l \leq 10.4^{\circ}$ and $-10.3^{\circ} \leq b \leq 5.1^{\circ}$) and some areas of the inner disk ($294.7^{\circ} \leq l \leq 350^{\circ}$ and $-2.25^{\circ} \leq b \leq 2.25^{\circ}$). 
The survey was conducted using the VIRCAM camera on the 4.1 m VISTA telescope located in Cerro Paranal Observatory, in Chile. It contains multi-band photometric ($Z, Y, J, H$, $K_{\mathrm{s}}$) information for 348 fields, with 196 fields in the Bulge and 152 fields in the southern disk, each single field covering 1.5$^\circ$ $\times$ 1.2$^\circ$ of sky (referred to as a ``tile"). Figure \ref{fig:2.1} shows the Bulge region and tile numbers in the VVV survey. 

% \begin{wrapfigure}{l}{0.45\textwidth}
%     \centering
%     \includegraphics[scale=0.4]{Figs/ms2025-0523fig1.png}
%     \caption{Schematic diagram of the Bulge region in the VVV survey, with numbers indicating the tile IDs.}
%     \label{fig:2.1}
% \end{wrapfigure}

The official data processing pipeline of VVV is the Cambridge Astronomical Survey Unit (CASU; \citealp{emerson_vista_2004}, \citealp{irwin_vista_2004}, \citealp{hambly_vista_2004}). It is capable of reducing and combining stacked pawprints and tiles, performing astrometry, and calibrating VVV observations. CASU also provides aperture photometry catalogs for every tile. However, for the VVV, PSF photometry is more robust and accurate than aperture photometry in the highly crowded fields. \cite{alonso-garcia_milky_2018} published Dophot-based PSF-photometry that is sufficiently deep to sample the Main Sequence turnoff (MS-TO) of the older population present in the Bulge for most fields. However, they only tested artificial stars in three representative fields, focusing on completeness.

\cite{surot_mapping_2019_2} used DAOPHOT/ALLFRAME to obtain the PSF-fitting photometry catalogue of VVV and published it named MW-BULGE-PSFPHOT in the ESO Archive Science Portal. They not only calculated the completeness of each tile but also mapped the extinction of the sky area covered by these catalogs. 
Using  Red Giant Branch (RGB) and RC stars they derived a high-resolution $E(J–K_{\mathrm{s}})$ reddening map \citep{surot_mapping_2020}.

\begin{figure}[!htb]
\centering  % 图片全局居中
\begin{subfigure}{\textwidth}
\centering
\includegraphics[scale=0.45]{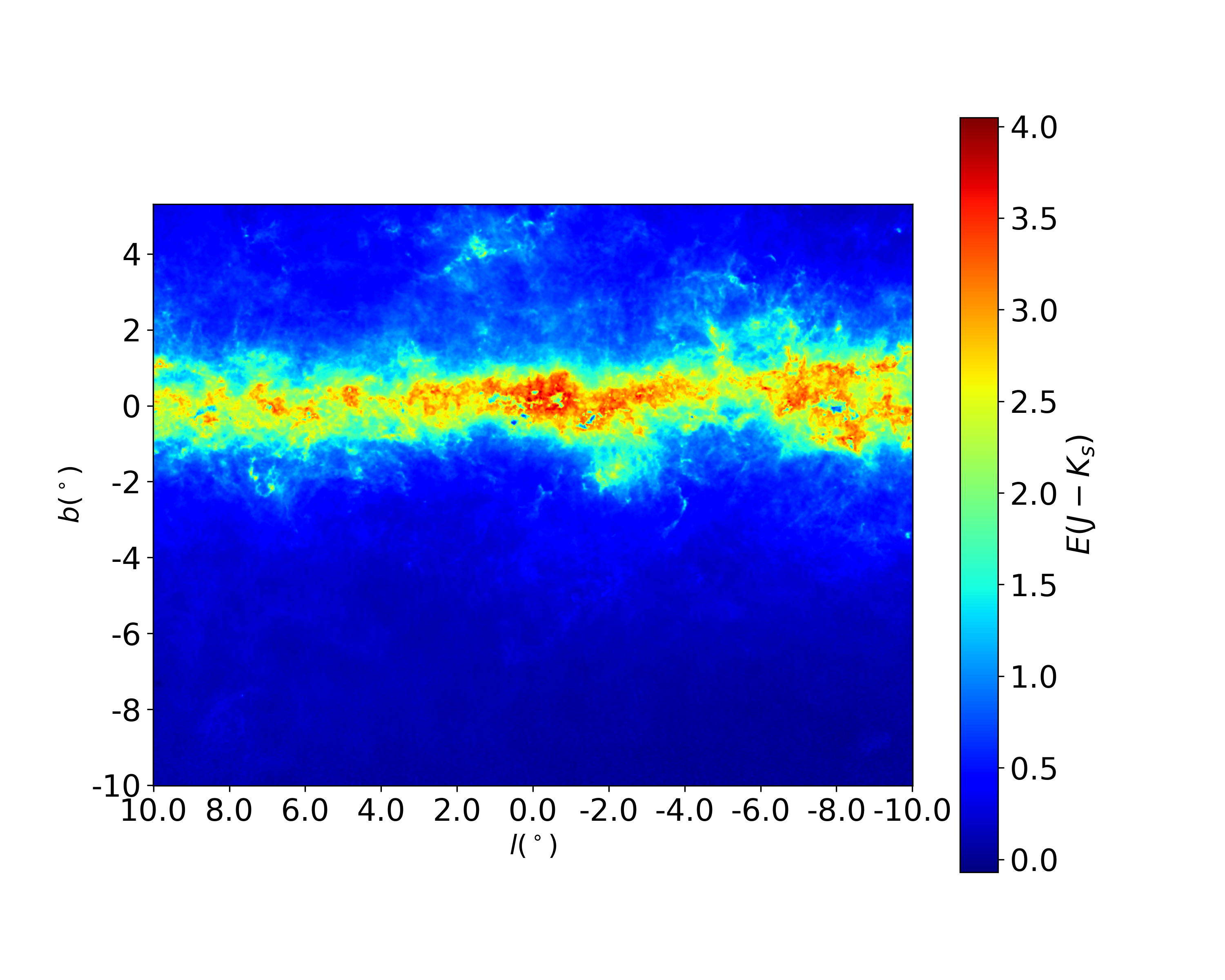}
\caption{}
\label{fig:2.2}
\end{subfigure}
\vfill
\begin{subfigure}{\textwidth}
\centering
\includegraphics[scale=0.45]{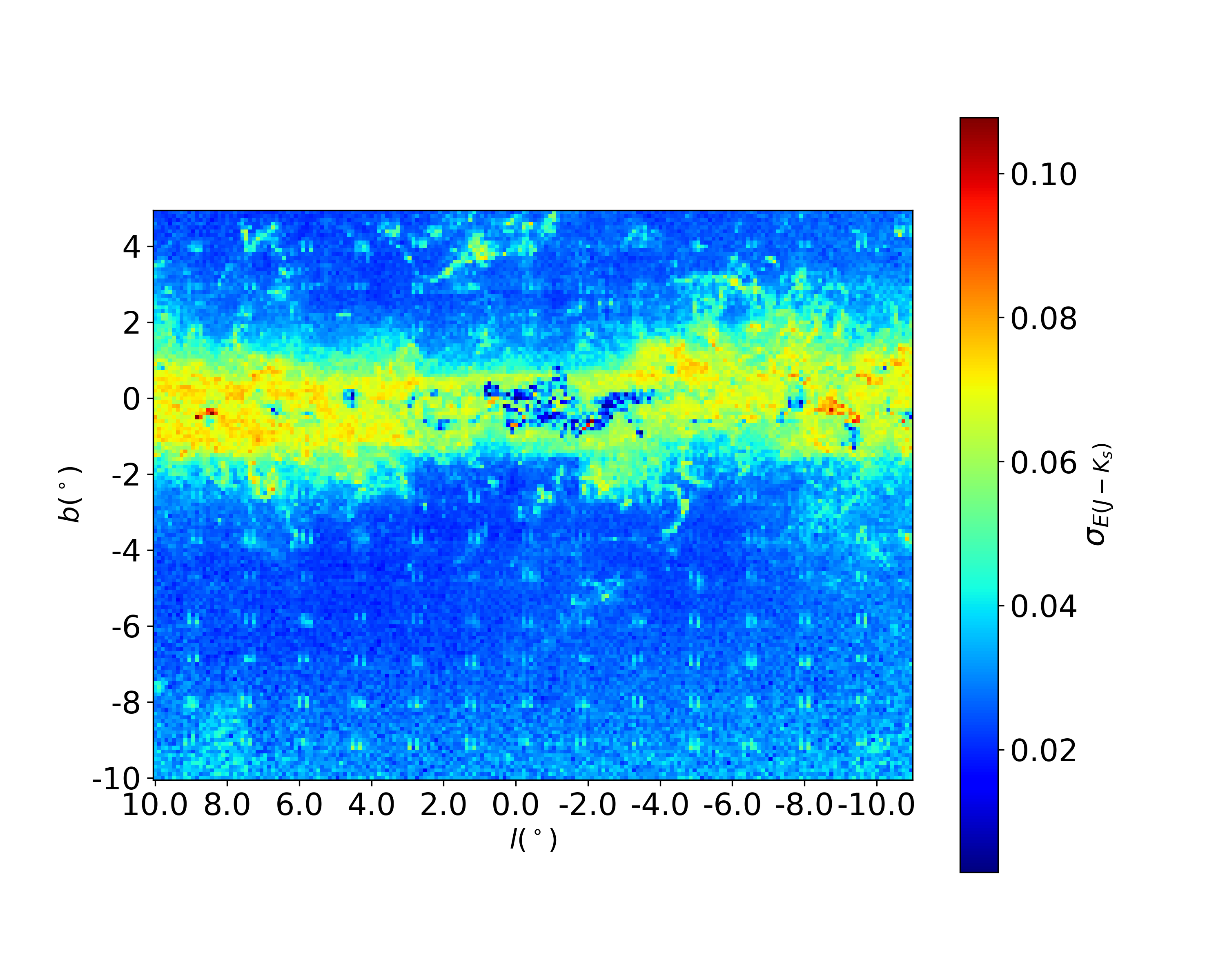}
\caption{}
\label{fig:2.2-1}
\end{subfigure}
\caption{(a) Reddening map in the Bulge area from \cite{simion_parametric_2017}, sampled at 1' $\times$ 1'; (b) Reddening error map.}
\label{fig:Extinction}
\end{figure}

However, since the sample we use for age determination consists of red giant  (RG) stars (see Section \ref{sec:IF} for details), we adopt the reddening results from Figure 1 in \cite{simion_parametric_2017} instead of the results in \cite{surot_mapping_2020} to avoid ambiguity. Similar to \cite{gonzalez_inner_2011}, \cite{simion_parametric_2017} derived a high-resolution (1' $\times$ 1' sampling) reddening map by comparing mean RC stars ($12 < m_{K_{\mathrm{s}}} < 14~\text{mag}$) colors across VVV fields. They used the RC giants as standard candles,  and the resulting reddening is converted to extinction via the \cite{nishiyama_interstellar_2009} law before being interpolated to correct individual stars. The mean color of RC giants exhibits minimal dependence on parameters such as metallicity and age, making them a more reliable tool for measuring extinction \citep{gonzalez_reddening_2012}. Figure \ref{fig:Extinction} presents the reddening results and associated uncertainties from \cite{simion_parametric_2017}. Apart from the heavily extinguished central region, the mean error reaches a maximum of approximately 0.06.

In summary, we use the PSF-fitting photometry from 196 tiles (Bulge area) of the MW-BULGE-PSFPHOT catalogs and the reddening map covering the Bulge area from \cite{simion_parametric_2017} to obtain the stellar average age.

\section{Method} \label{sec:Mth}
In this section, we take the b393 tile ($l\sim 5.35^\circ$, $b\sim 4.49^\circ$) as a case study to demonstrate how to obtain the stellar age distribution of a single tile.
\subsection{Source selection}
\label{sec:SS}
We use Gaia DR3 to eliminate foreground stars according to \cite{marchetti_blanco_2022}. We first cross-match Gaia DR3 with the stellar positions (RAJ2000 and DECJ2000) in the dataset using a $1''$ radius. Afterward, we compute the parallax zeros for sources simultaneously in VVV and Gaia using the code of \cite{lindegren_gaia_2021}. We define the foreground stars as:

\begin{equation}  
\label{eqn:1}
\begin{aligned}
RUWE<1.4,\\
\omega-\omega_0>0.2~\text{mas},\\
\text{and } \sigma_{\omega}/(\omega-\omega_0)<0.2,
\end{aligned}
\end{equation}
where $\omega_0$ and $\sigma_{\omega}$ are the parallax zero-point and parallax uncertainty, respectively. The Renormalized Unit Weight Error (RUWE, \citealp{lindegren_gaia_2021}) is Gaia’s standard metric for assessing the quality of a single-star astrometric solution. Following Eq. (6) and Eq. (7) in \cite{marchetti_blanco_2022}, the first part of Eq. (\ref{eqn:1}) indicates a distance smaller than 5 kpc, while the second part considers the uncertainty of the Gaia parallax. These criteria ensure that the remaining stars are either too distant or have Gaia parallax measurements with insufficient accuracy, making it impossible to determine their distances through parallax inversion (e.g., \citealp{bailer-jones_estimating_2015}).

\begin{figure}[!htb]
\centering
\includegraphics[scale=0.2]{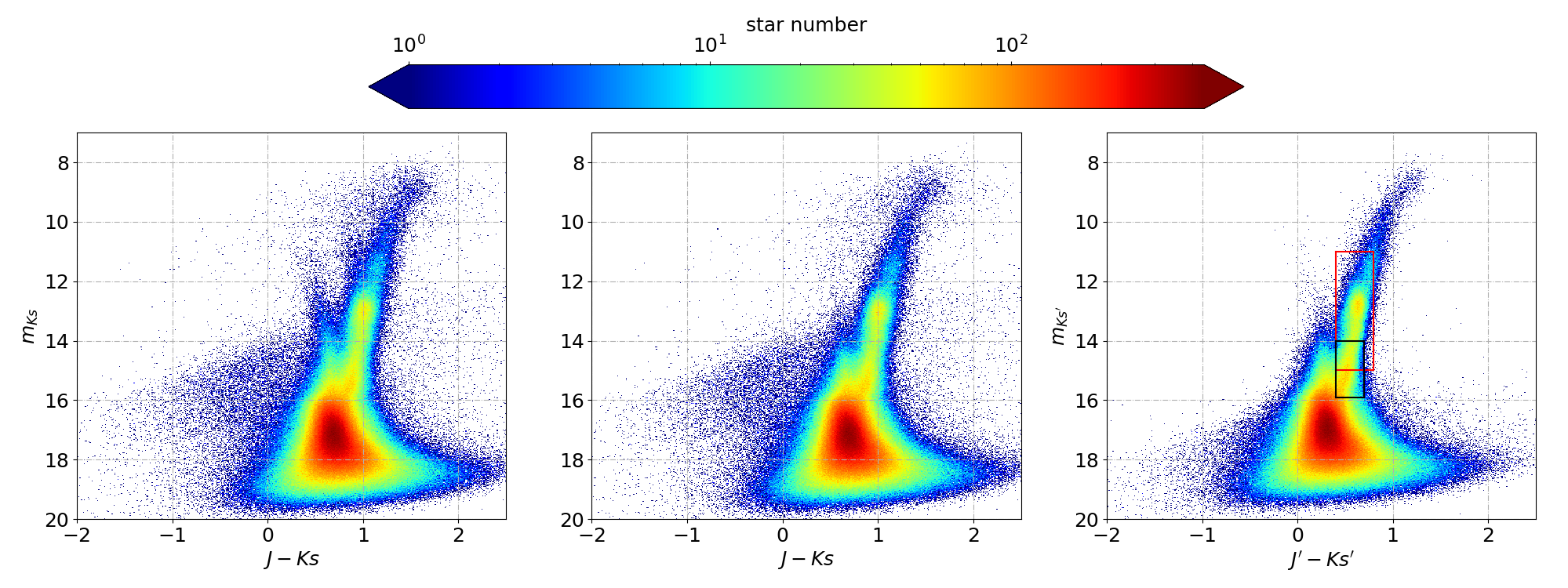}
\caption{
The CMD of tile b393. \textit{Left panel}: all the sources in b393. \textit{Middle panel}: the sources remaining after discarding the foreground stars. \textit{Right panel}: the remaining sources after removing the foreground stars and those with poor photometric quality. $J'$ and $K_{\mathrm{s}}'$ are the apparent magnitudes after extinction correction. The red box contains samples used for calculating the average distance, and the black box covers the samples used for fitting the average age.}
\label{fig:3.1}
\end{figure}

After discarding the foreground stars, we also remove sources with poor photometry based on the parameters (\textit{sharpness} and $\chi^2$) used by \cite{surot_mapping_2019_2} to assess the PSF photometric quality. We retain sources with $\chi^2 < 3$ and \textit{sharpness} $< 3$. Figure \ref{fig:3.1} shows the color-magnitude diagram (CMD) of b393, where the vertical coordinate represents the apparent magnitude in the $K_{\mathrm{s}}$ band and the color bar represents the number of stars. 
The left panel contains all the sources in b393. The middle panel shows the sources remaining after removing the foreground stars. The right panel shows the sources left over after discarding both foreground stars and those with poor photometric quality. $J'$ and $K_{\mathrm{s}}'$ represent the apparent magnitudes after extinction correction (the extinction map from \citealp{simion_parametric_2017}). After removing the bright and foreground stars, only one of the two protrusions remains in the region where $m_{K_{\mathrm{s}}'} < 14$ mag. Furthermore, after discarding sources with poor photometric quality, the CMD becomes cleaner.

\subsection{Average distance}
\label{sec:AD}
\begin{figure}[!htb]
\centering % 图片全局居中
\begin{subfigure}{0.45\textwidth}
\includegraphics[scale=0.32]{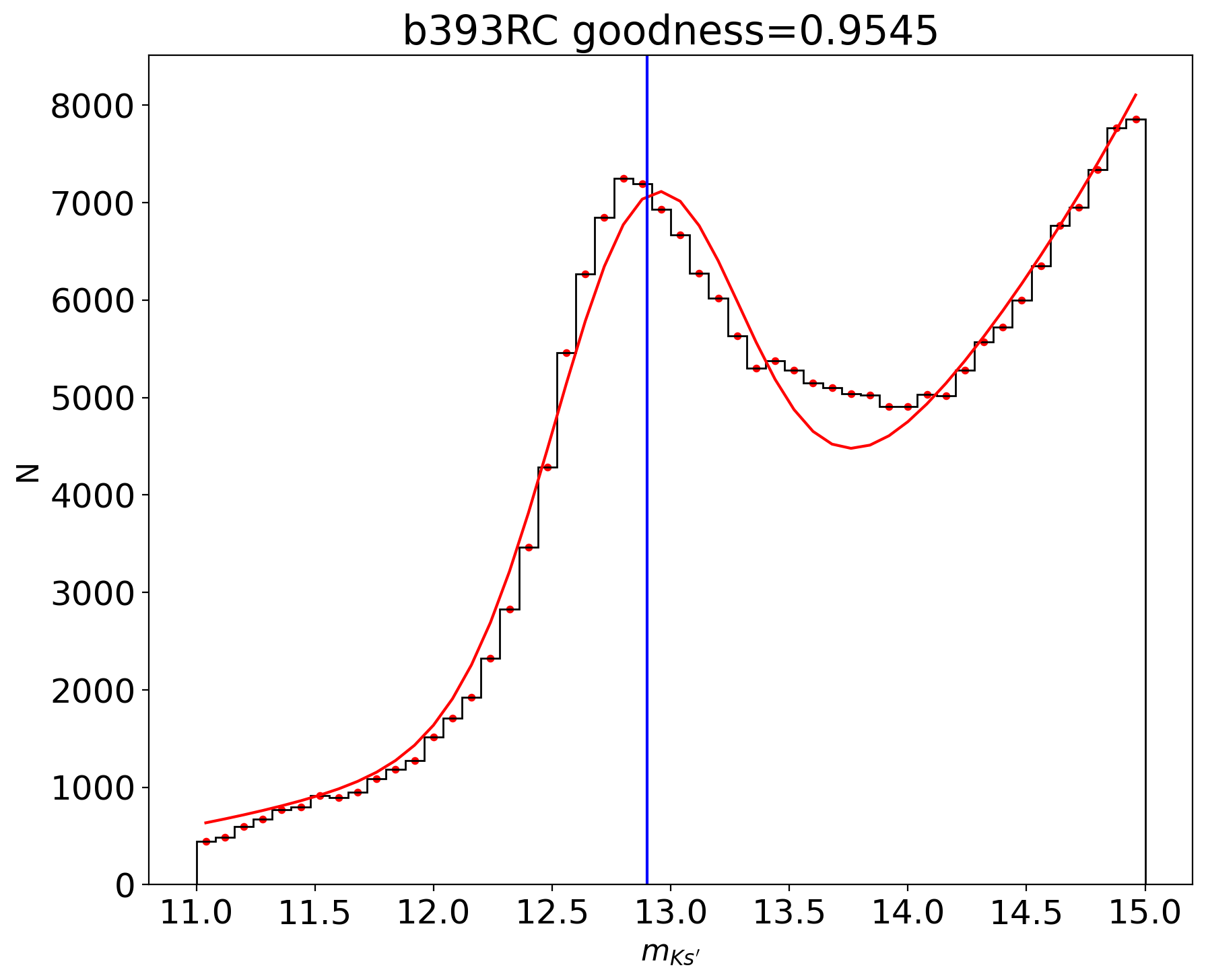}
\caption{}
\label{fig:3.2}
\end{subfigure}
\hfill
\begin{subfigure}{0.45\textwidth}
\includegraphics[scale=0.32]{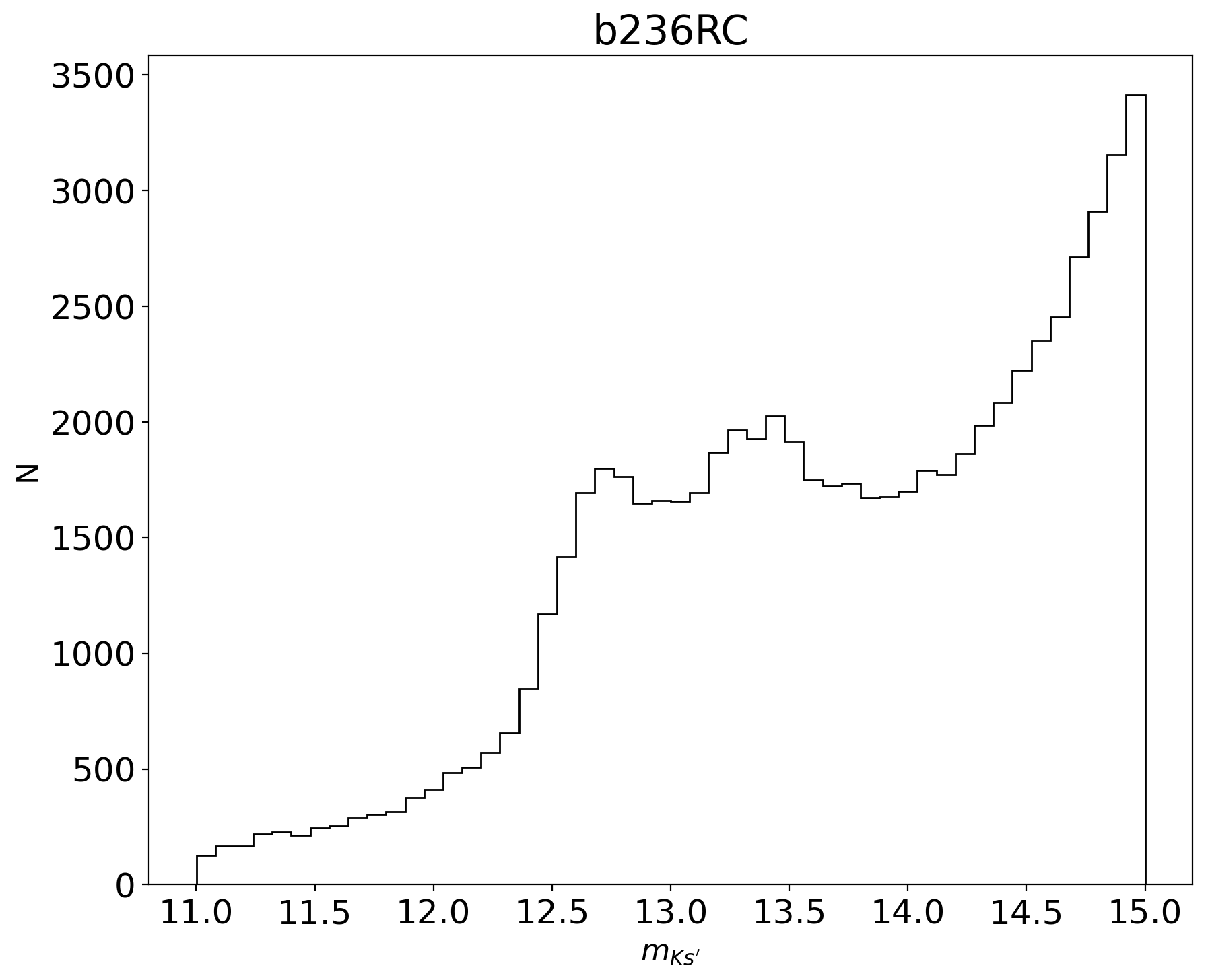}
\caption{}
\label{fig:3.2-1}
\end{subfigure}
\caption{(a) Apparent magnitude ($K_{\rm s}$ band) distribution of the RC stars in b393 tile ($l \sim 5.35^\circ$, $b \sim 4.49^\circ$); (b) An example of the bimodal distribution of the RC stars, b236 tile ($l \sim 1.05^\circ$, $b \sim -7.47^\circ$).}
\label{fig:mb}
\end{figure}

Because the parallax accuracy and completeness of the Galactic Bulge ($\sim$ 8 kpc) in Gaia are not high enough, as well as the selection criterion based on the equation (\ref{eqn:1}), the average distance computed directly from parallax will be affected by a large selection effect and uncertainty. Therefore, we rely on RC stars to calculate the average distance of Bulge stars. The RC stellar population is widely adopted as a standard candle, owing to its narrow luminosity function in the near-infrared. Its calibrated absolute magnitude shows little dependence on metallicity, enabling precise distance estimates to galactic and nearby extragalactic targets \citep{Paczyński_2016,girardi_red_2016}.
According to the right panel of Figure \ref{fig:3.1}, we select the stars in 11 $< m_{K_{\mathrm{s}}'} <$ 15 and 0.4 $< J'-K_{\mathrm{s}}' <$ 0.8 represents the range of RC stars (the red box). After that, we obtain the mean value of the apparent magnitude of the RC stars based on their distribution (histogram in Figure \ref{fig:3.2}). First, we divide the stars in the red box into 50 groups according to $m_{K_{\mathrm{s}}'}$. Then we fit the distribution of the apparent magnitude with a function (Eq. (\ref{eqn:3})) that is a mixture of Gaussian (standing for the distribution of RC stars) and a power-law (standing for contaminated RGB stars) to obtain its mean value:
\begin{equation}
\label{eqn:3}
\begin{aligned}
N(m) = \frac{N_{RC}}{\sigma_{RC}\sqrt{2\pi}}\exp\Bigg[-\frac{(m_{RC}-m)^2}{2\sigma^2_{RC}}\Bigg]+Am^B.
\end{aligned}
\end{equation}

We use the least squares function (the least\_squares package in Python's scipy library) to fit and calculate the goodness of fit. The blue line in Figure \ref{fig:3.2} is the average $m_{K_{\mathrm{s}}'}$ of b393 tile in the fit and the red line is the fitting result function form.

After obtaining the average $m_{K_{\mathrm{s}}'}$, we use the following equation to derive the distance and the distance modulus:
\begin{equation}
\label{eqn:2}
\begin{aligned}
m_{K_{\mathrm{s}}’}-M_{K_{\mathrm{s}}}=5\log(d)-5,
\end{aligned}
\end{equation}
where $M_{K_{\mathrm{s}}}=-1.6$ (\citealp{hawkins_red_2017}). 
It is important to note that when calculating the average distance of the single tile, we assume that all Bulge member stars in b393 tile have the same average distance.

For some other fields, we find that the magnitude distribution has a bimodal structure, such as tile b236 (Figure \ref{fig:3.2-1}). They may be related to the X-shape of the Galactic Bulge \citep{mcwilliam_two_2010, saito_mapping_2011} or to a superposition of a main+secondary RC stars \citep{lopez-corredoira_distribution_2019}. 
We remove these regions from the following calculation of the average age. So we throw away 46 tiles, leaving 150 tiles. 

\subsection{Isochrone fitting}
\label{sec:IF}
\begin{figure}[ht!]
\centering
\includegraphics[scale=0.34]{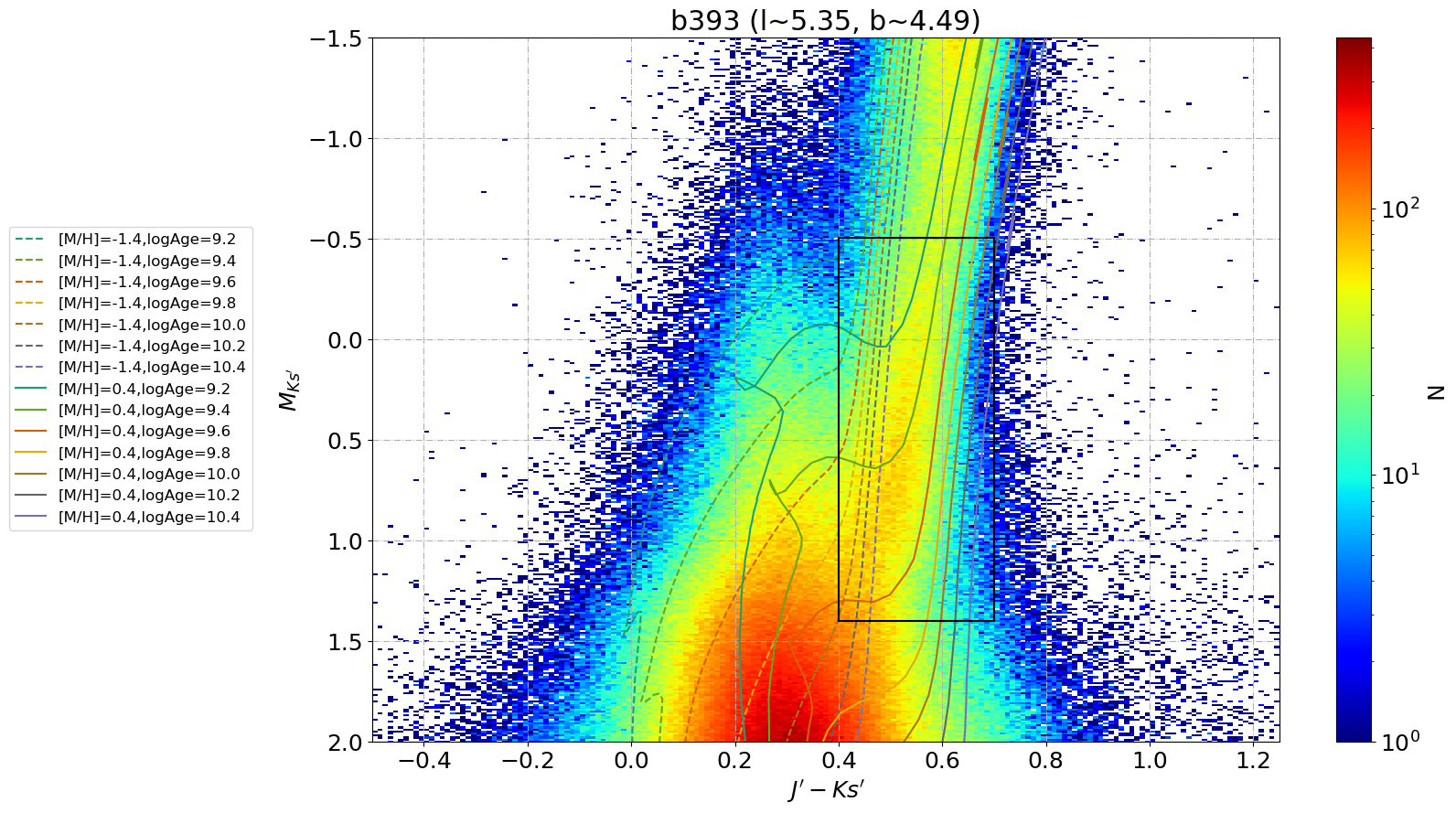}
\caption{The CMD for b393 tile. Similar to the right panel of Figure \ref{fig:3.1} but with absolute magnitude as the ordinate. The black box outlines the sample used for age fitting. The metallicities and logAge of the isochrones, shown in different colors, are provided in the legend. To better illustrate, we select only two sets of isochrones with different [M/H]. The isochrone for [M/H] = -1.4 is represented by the dashed line, and [M/H] = 0.4 is depicted by the solid line. The same age share same colors.
}
\label{fig:3.3}
\end{figure}

In this section, we determine the age distribution of stars by fitting RG stars with isochrones in CMD. 
We select the stars with $J'-K_{\mathrm{s}}'\in[0.4, 0.7]$, $m_{K_{\mathrm{s}}'}>14$ mag and $M_{K_{\mathrm{s}}}<1.4$ mag as our sample (the black box in the right panel in Figure. \ref{fig:3.1}). The primary reasons for selecting this subset of stars as the sample for age fitting are as follows: 
a) As described in Section \ref{sec:Data}, the extinction results we use are derived from stars within the magnitude range 12 mag $<m_{K_{\mathrm{s}}}<$ 14 mag. Therefore, all stars chosen for age fitting are fainter than 14 mag. 
b) Considering stellar photometric errors and completeness, we ultimately select RG stars located along the ridgeline of the CMD as tracers for age determination. The theoretical isochrones are obtained from the PARSEC v1.25+COLIBRI S37+COLIBRI S35+COLIBRI PR16 evolutionary tracks \citep{marigo_evolution_2013,chen_improving_2014,tang_new_2014,chen_parsec_2015,rosenfield_evolution_2016,pastorelli_constraining_2019,pastorelli_constraining_2020} with magnitudes in the VISTA bands (Vegamag). We use a total of 49 isochrones with various ages and metallicity. The logarithmic value of age (logAge) $\in$ [9.2,10.4] with a step of 0.2 (7 sampling points) and the metallicity ([M/H]) $\in$ [-1.4,0.4] with the step of 0.3 (also 7 sampling points). Note that all of the following descriptions related to age and metallicity are essentially about logAge and [M/H]. Taking the b393 tile as an example, Figure \ref{fig:3.3} presents a zoomed-in view of the sample used for age fitting in the CMD, with overlaid isochrones. The black box outlines the sample used for age fitting. For clarity, not all isochrones employed in the fitting process are displayed. We now provide a detailed description of the procedure for calculating the mean age using isochrones. It is important to emphasize that our approach does not involve selecting a single best-fitting isochrone from the grid. Instead, we reconstruct the distribution of the stellar color ($J'-K_{\mathrm{s}}'$) by performing a weighted average over the isochrones, incorporating observational errors, and then derive the age distribution of the sample based on the discrepancy between the reconstructed and observed colors. 

We first divide the sample into nine equally spaced bins in M$_{K_{\mathrm{s}}}$.
In each M$_{K_{\mathrm{s}}}$ bin, we can obtain the cumulative distribution function (CDF) about $J'-K_{\mathrm{s}}'$ (CDF$_{obs\_{M_{K_{\mathrm{s}}}}}$), composed of star populations of different ages and metallicity. 

For the theoretical isochrones, we interpolate the M$_{K_{\mathrm{s}}}$ to be consistent with the observed M$_{K_{\mathrm{s}}}$. 
The CDF consisting of isochrones in an M$_{K_{\mathrm{s}}}$ bin (CDF$_{M_{K_{\mathrm{s}}}}$) is: 

\begin{equation}
\label{eqn:4-1}
\begin{split}
CDF_{M_{K_{\mathrm{s}}}} = \int PDF_{M_{K_{\mathrm{s}}}}\mathrm{d}(J'-K_{\mathrm{s}}'),
\end{split}
\end{equation}
where $PDF_{M_{K_{\mathrm{s}}}}$ is the probability density distribution (PDF) of $M_{K_{\mathrm{s}}}$. It can be represented as
\begin{equation}
\label{eqn:4-2}
\begin{split}
PDF_{M_{K_{\mathrm{s}}}} = \int\int f'(a,m)P(J'-K_{\mathrm{s}}'|a,m)\Delta a \Delta m,
\end{split}
\end{equation}
where $a$ and $m$ represent logAge and [M/H], respectively.
\begin{equation}
\label{eqn:4-3}
\begin{split}
f'(a,m) = f(a|m)f'(m),
\end{split}
\end{equation}
where $f(a|m)$ is the distribution of $a$ given $m$. So
\begin{equation}
\label{eqn:4-3-1}
\begin{split}
f(a|m) = \frac{f(a,m)}{f(m)},
\end{split}
\end{equation}
where f(a,m) is the joint distribution of $a$ and $m$. And f(m) is the marginal distribution, $f(m)=\int f(a,m)  \, da$.

Due to the degeneracy relationship between age and metallicity, we select results from spectroscopy as the prior $f'(m)$ to impose constraints on metallicity, thereby reducing the influence of the age-metallicity relationship on the subsequently calculated age results. $f'(m)$ in Eq. (\ref{eqn:4-3}) is the distribution function of metallicity in different latitude ranges, described in a multi-Gaussian form, as provided by \cite{rojas-arriagada_how_2020}.
According to \cite{rojas-arriagada_how_2020}, it consists of Metal-poor, Metal-intermediate, and Metal-rich components, all of which can be represented by the Gaussian distribution:
\begin{equation}
\label{eqn:8}
\begin{aligned}
f'(m)=\sum_{i=1}^3 w_iN(\mu_i,\sigma_i),
\end{aligned}
\end{equation}
where $w_i$ is the weight of each Gaussian. The values of $w$, $\mu$, and $\sigma$ for different latitudes ($b$) are from Table 2 of \cite{rojas-arriagada_how_2020}. Their [M/H] distribution function was obtained by fitting the spectroscopic data from APOGEE using Gaussian mixture modeling.

And then, we convert Eq. (\ref{eqn:4-2}) into discrete form:

\begin{equation}
\label{eqn:4-4}
\begin{split}
PDF_{M_{K_{\mathrm{s}}}}&=\sum_{i=1}^7\sum_{j=1}^7 f'(a_i,m_j)P(J'-K_{\mathrm{s}}'|a_i,m_j)\Delta a\Delta m \\
&=\sum_{i=1}^7\sum_{j=1}^7f(a_i|m_j)f(m_j)P(J'-K_{\mathrm{s}}'|a_i,m_j)\Delta a\Delta m\\
&=\sum_{i=1}^7\sum_{j=1}^7\frac{f(a_i,m_j)}{f(m_j)}f'(m_j)P(J'-K_{\mathrm{s}}'|a_i,m_j)\Delta a\Delta m,
\end{split}
\end{equation}
where $a_i$ and $m_j$ are the sampling point for logAge and [M/H] of isochrone, respectively. We assume that the stellar populations obey a 2D Gaussian distribution in the (logAge, [M/H]) plane:
\begin{equation}
\label{eqn:4-5}
\begin{aligned}
f(a_i, m_j) = & \left( \frac{1}{2\pi \sigma_{\log Age} \sigma_{[M/H]} \sqrt{1 - \rho^2}} \right) \\
& \times \exp \left[ - \frac{1}{2(1 - \rho^2)} \left( \frac{(a_i - \mu_{\log Age})^2}{\sigma_{\log Age}^2} \right. \right. \\
& \left. \left. - \frac{2\rho(a_i - \mu_{\log Age})(m_j - \mu_{[M/H]})}{\sigma_{\log Age} \sigma_{[M/H]}} + \frac{(m_j - \mu_{[M/H]})^2}{\sigma_{[M/H]}^2} \right) \right],
\end{aligned}
\end{equation}
where $\mu$ and $\sigma$ are the average and standard deviation of logAge and [M/H], respectively. $\rho$ is the correlation coefficient of logAge and [M/H]. In addition, we assume that 
\begin{equation}
\label{eqn:4-6}
\begin{split} 
P(J'-K_{\mathrm{s}}’|a_i,m_j)\sim T(\overline{C},\sigma), \\
\end{split}
\end{equation}
where $T(\overline{C},\sigma)$ is a truncated Gaussian distribution on $J'-K_{\mathrm{s}}’$ $\in$ [0.4,0.7]. We interpolate each isochrone of each bin to make the number of isochrone sampling points 100, after which we calculate the mean of their $J'-K_{\mathrm{s}}’$ as $\overline{C}$. The $\sigma$ is the error of the photometry and extinction from the sample. Actually, we use the distribution of these isochrones ($P(J'-K_{\mathrm{s}}'|a_i,m_j)$) to reconstruct a new PDF ($PDF_{M_{K_{\mathrm{s}}}}$). $f'(a_i,m_j)$ in Eq. (\ref{eqn:4-4}) is the ratio of each isochrone, so we normalize it to ensure that the sum of the ratios is equal to 1.

Finally, we use the Kolmogorov--Smirnov test to obtain the distance between $\mathrm{CDF}_{\mathrm{obs}\_M_{K_{\mathrm{s}}}}$ and $\mathrm{CDF}_{M_{K_{\mathrm{s}}}}$ in each $M_{K_{\mathrm{s}}}$ bin, interpolating both CDFs onto 10 000 points:
\begin{equation}
\label{eqn:5}
\begin{aligned}
D_{color} = Max(|CDF_{obs\_{M_{K_{\mathrm{s}}}}}-CDF_{M_{K_{\mathrm{s}}}}|).
\end{aligned}
\end{equation}
To get a stable fitting result, we consider the completeness of the sample. We calculate the weighted average of the distances ($\bar{D}_{color}$) for all bins, weighted as the average completeness (for detailed numerical values, see Section \ref{sec:Com&SE}) of the samples in different M$_{K_{\mathrm{s}}}$ bins. 
And then, we also compute the distance ($D_{[M/H]}$) between the CDF of the model-fitted [M/H] distribution, $f(m)$, and that of the spectroscopic prior, $f'(m)$ (Eq. (\ref{eqn:8})).
After that, we define the following likelihood function:
\begin{equation}
\label{eqn:9}
\begin{aligned}
L= \bar{D}_{color}+s\times D_{[m/H]},
\end{aligned}
\end{equation}
where s is a shrinkage factor that ensures $(\bar{D}_{color})$ and $(D_{[m/H]})$ are the same order of magnitude. The free parameters are $\mu_{logAge}, \mu_{[M/H]}, \sigma_{logAge}, \sigma_{[M/H]}$ and $\rho$ in Eq. (\ref{eqn:4-5}). We use the MCMC sampling to find the best convergence results of free parameters that minimise $ln L$. We set up 100 samplers in the MCMC process, with burn-in and formal sampling steps of 1000 and 2000, respectively. 

\section{Results} \label{sec:Re}
Below, we present the parameter constraints derived from the MCMC sampling described in the previous section.  We first focus on the single VVV tile b393 to validate the method and establish baseline values for the five free parameters. We then extend the analysis to all remaining sky regions, presenting the average age and investigating its variation with Galactic latitude $|b|$.

\subsection{Single tile}
\label{sec:SF}

\begin{figure}[ht!]
\centering
\includegraphics[scale=0.5]{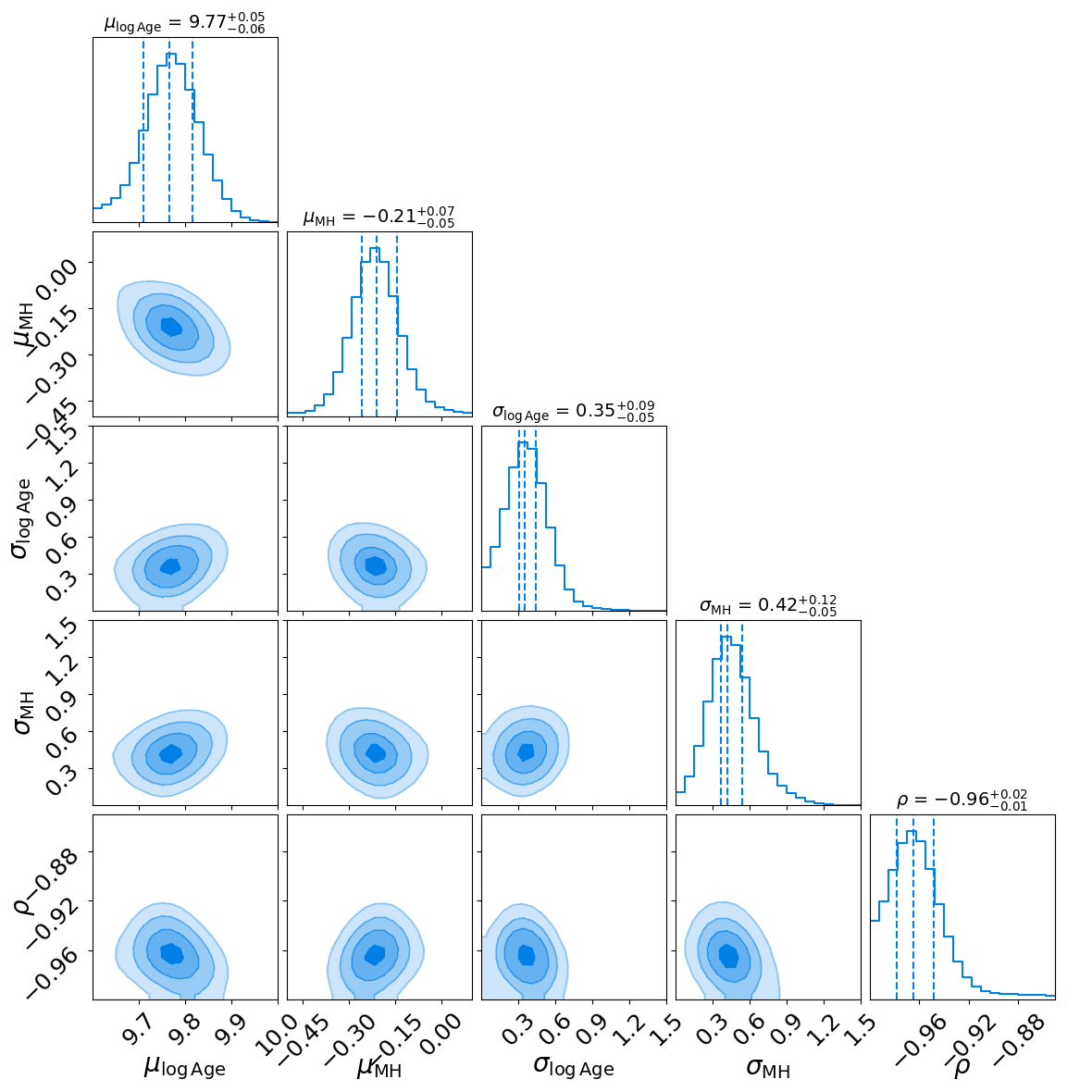}
\caption{Corner diagram of MCMC parameters. In the 1D histograms, dashed lines indicate the 0.16, 0.5, and 0.84 quantiles. The best-fit results correspond to the values at the 0.5 quantile.
\label{fig:4.1-1}}
\end{figure}

Figure \ref{fig:4.1-1} illustrates the MCMC results for the tile b393. It shows that the logAge follows a Gaussian distribution with an average ($\mu_{logAge}$) of 9.77 and a standard deviation ($\sigma_{logAge}$) of 0.35; the [M/H] follows a Gaussian distribution with an average ($\mu_{[M/H]}$) of -0.21 and a standard deviation ($\sigma_{[M/H]}$) of 0.42. The correlation coefficient is -0.96, showing that age and metallicity are strongly negatively correlated throughout the evolutionary stages of RG stars.  We define the intrinsic error (1$\sigma$ uncertainty) of each parameter as the mean of the upper and lower errors. The upper error is the difference between the 0.84 quantile and the 0.5 quantile, while the lower error is the difference between the 0.5 quantile and the 0.16 quantile. In the case of the b393 tile, the intrinsic error of the average logAge is $(0.05+0.06)/2\approx 0.055$. 

\begin{figure}[ht!]
\centering
\includegraphics[scale=0.15]{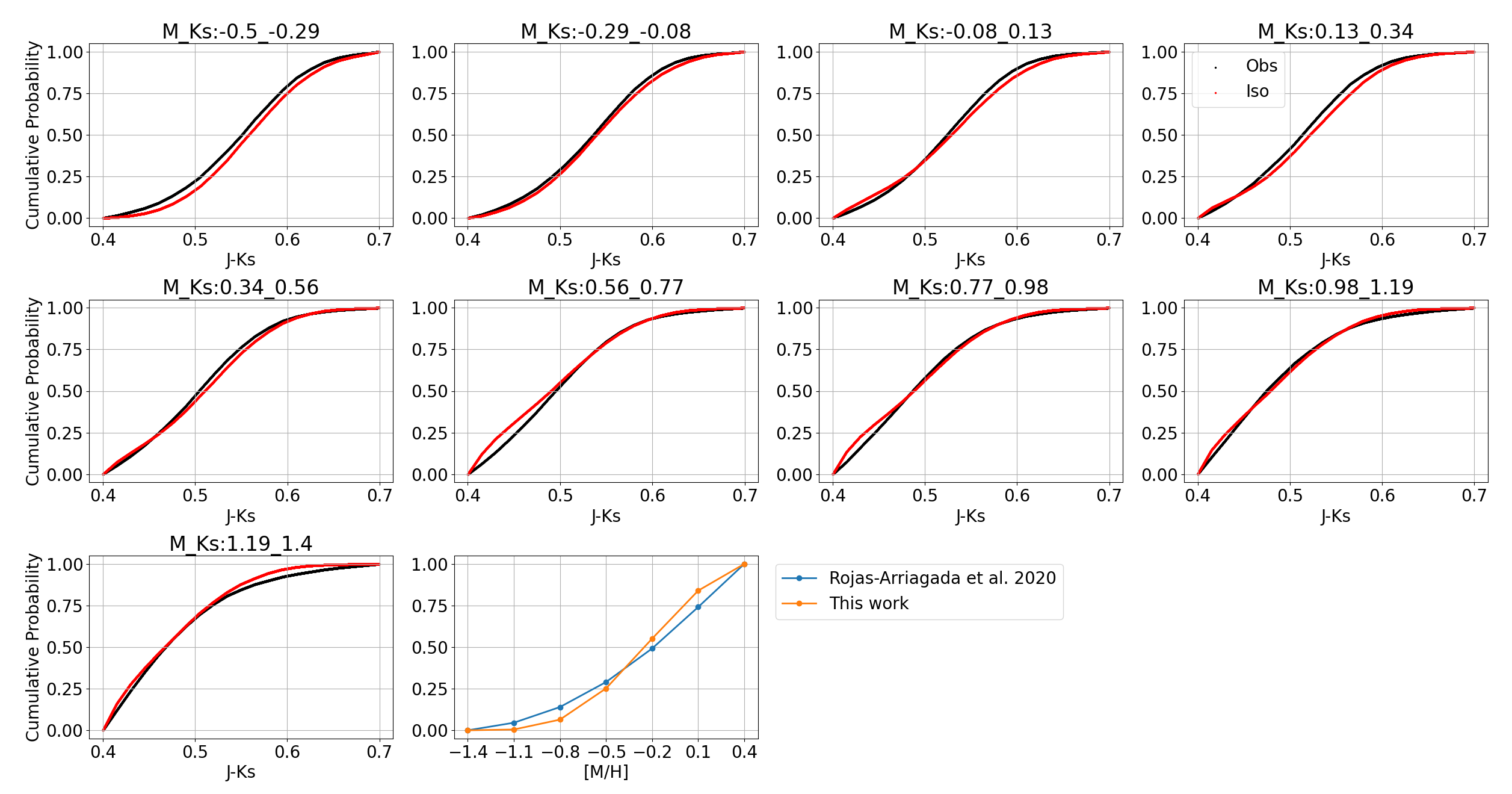}
\caption{Comparison of the fitted color CDF with the sample’s color CDF, and comparison of the fitted [M/H] CDF with the prior [M/H] CDF. The subheadings are absolute magnitude ranges for each group. The black dots represent the samples and the red dots represent the fitting results. The maximum difference between CDFs does not exceed 0.1 dex.
\label{fig:4.1-2}}
\end{figure}

Based on the results in Figure \ref{fig:4.1-1}, we compare the reconstructed CDF from the isochrone fitting with the CDF of the observed samples. We calculate the difference between these CDFs within each $M_{K_{\mathrm{s}}}$ bin. According to Figure \ref{fig:4.1-2}, the maximum difference is lower than 0.1. Likewise, the CDF obtained from the fitted [M/H] distribution differs from the one anchored to the prior [M/H] distribution (see Section~\ref{sec:IF} for details) by less than 0.1 dex. It should be noted that, in Figure \ref{fig:4.1-2}, the CDF of each color shown in the first nine panels is constructed from $10\,000$ interpolated sample points rather than from a fitted curve, whereas the CDF of [M/H] in the final panel uses the same sampling points as the isochrone grid (see Section~\ref{sec:IF} for details).

\subsection{Average logAge map of the Galactic Bulge}
\label{sec:GAL}

\begin{figure}[!htb]
% \centering  %图片全局居中
\begin{subfigure}{0.5\textwidth}
\centering
\includegraphics[scale=0.25]{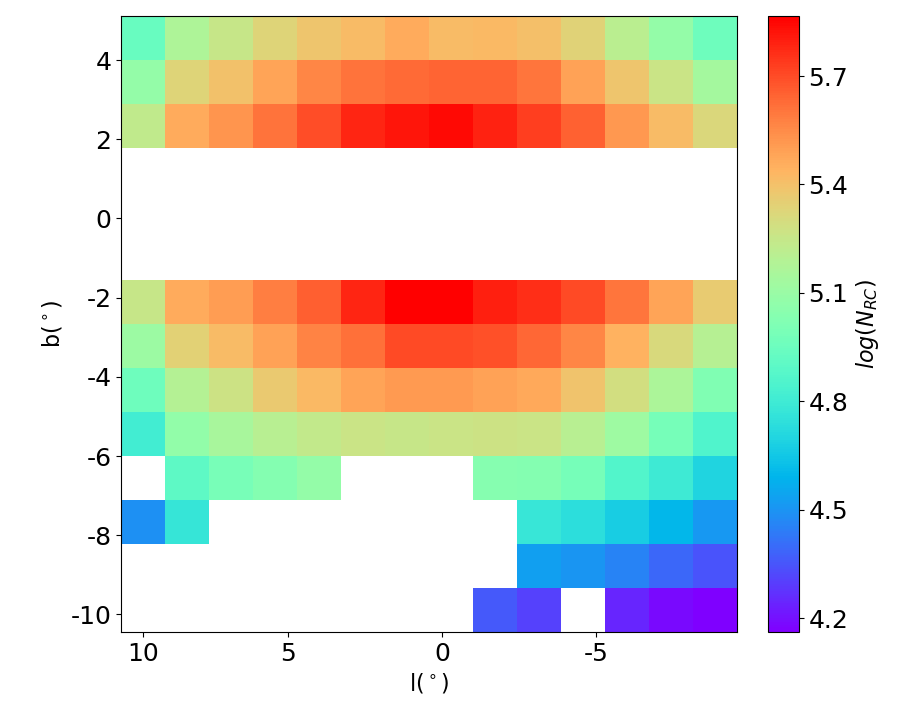}
\caption{}
\label{Fig.sub.1}
\end{subfigure}
\hfill
\begin{subfigure}{0.48\textwidth}
% \centering
\includegraphics[scale=0.25]{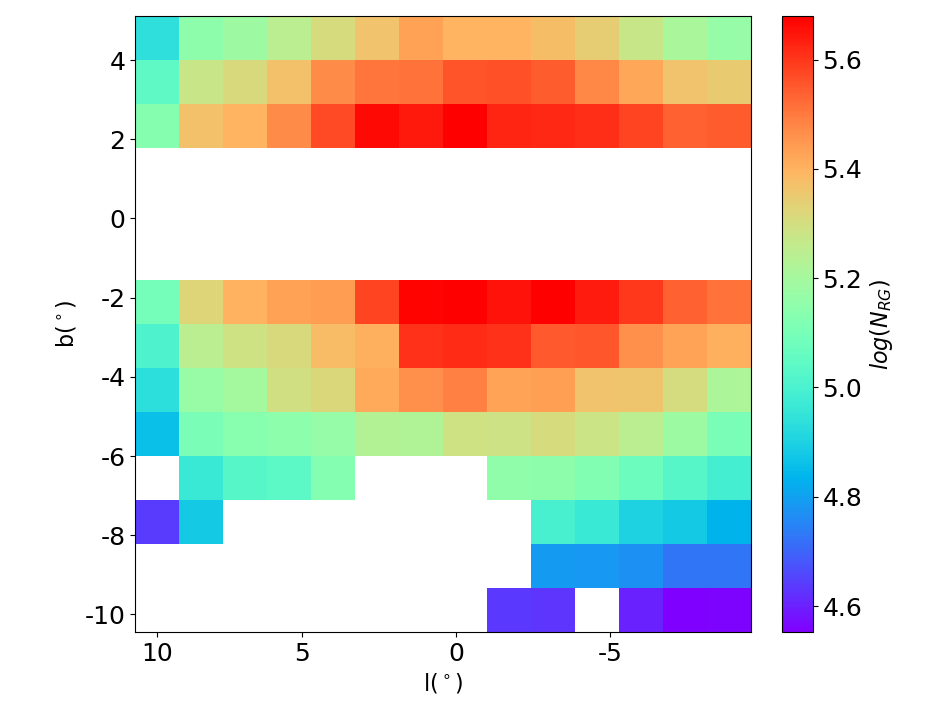}
\caption{}
\label{Fig.sub.2}
\end{subfigure}
\caption{(a) The number density distribution of RC stars. The color bar is logarithmic. (b) Same as (a), but for RG stars.
\label{fig:1}}
\end{figure}

After determining the average age of the individual tile (b393), we calculate the average age of the remaining tiles in the MW-BULGE-PSFPHOT catalogs. Because the central region suffers from severe extinction, we exclude this area (b313$–$b354 in Figure \ref{fig:2.1}) when constructing the global age map. We found that tiles with $|b|>8^{\circ}$ could not get a self-consistent set of results (MCMC results do not converge). We compared the number densities of RC and RG stars in all the tiles used to compute the average distance and average logAge. As seen in Figure \ref{fig:1}, the tiles with $|b| > 8^\circ$ have significantly smaller numbers of RC and RG stars than the other tiles. Blank areas in figure are thrown away tiles since their RC stellar magnitude distribution is bimodal (more details in Section \ref{sec:AD}).
To maintain the high precision of our age derivations, we focused on regions characterised by a single, well-defined RC peak. Since our methodology requires a stable average distance to accurately align theoretical isochrones with the observed CMD (more details see section \ref{sec:IF}), these bimodal regions were excluded from the current analysis. 
So we only show results for $|b|<8^{\circ}$. A total of 114 fields are finally used to construct the average age map of the Galactic bulge.

\begin{figure}[!htb]
\centering  %图片全局居中
\begin{subfigure}{0.5\textwidth}
\includegraphics[scale=0.32]{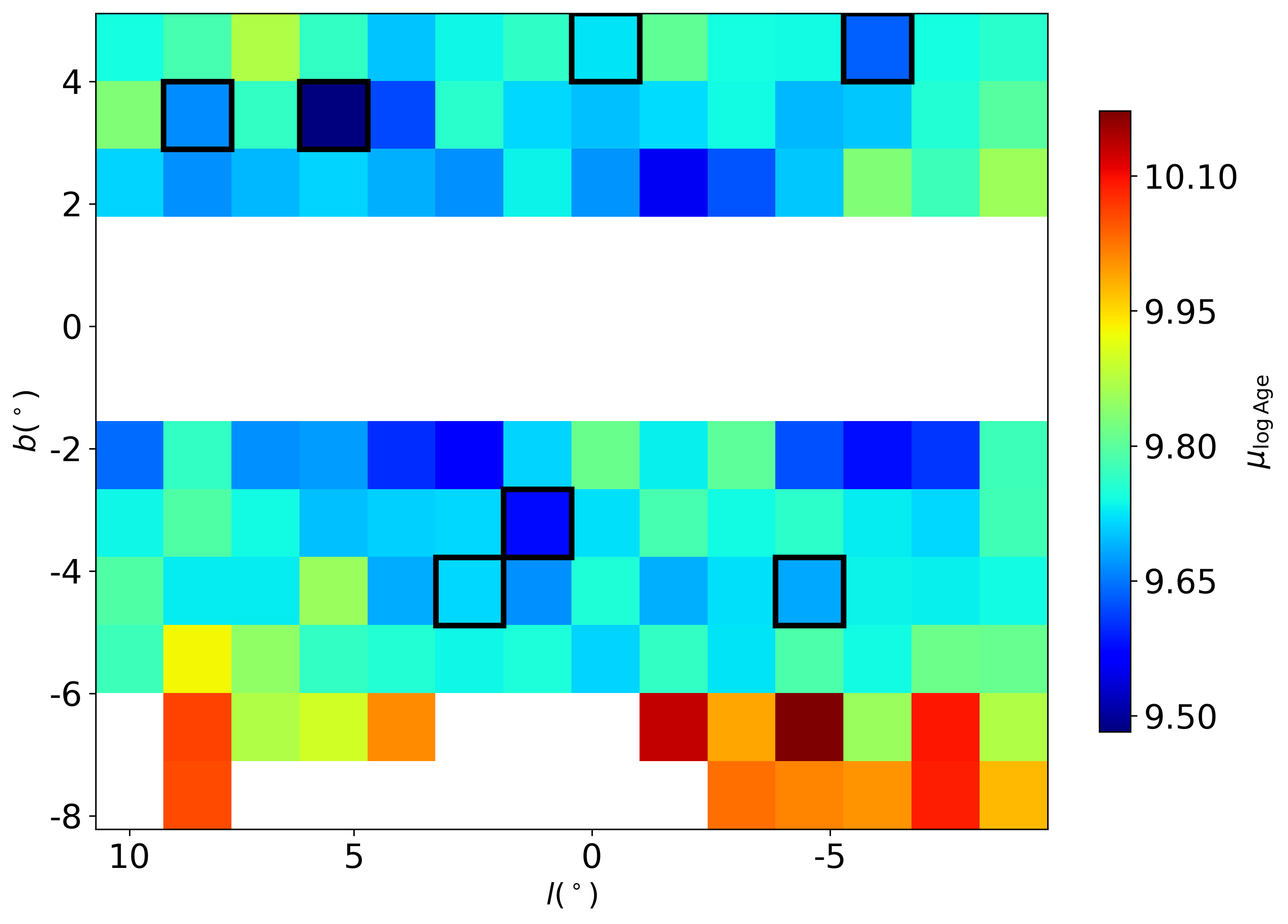}
\caption{}
\label{Fig.sub.3}
\end{subfigure}
\begin{subfigure}{0.48\textwidth}
\includegraphics[scale=0.32]{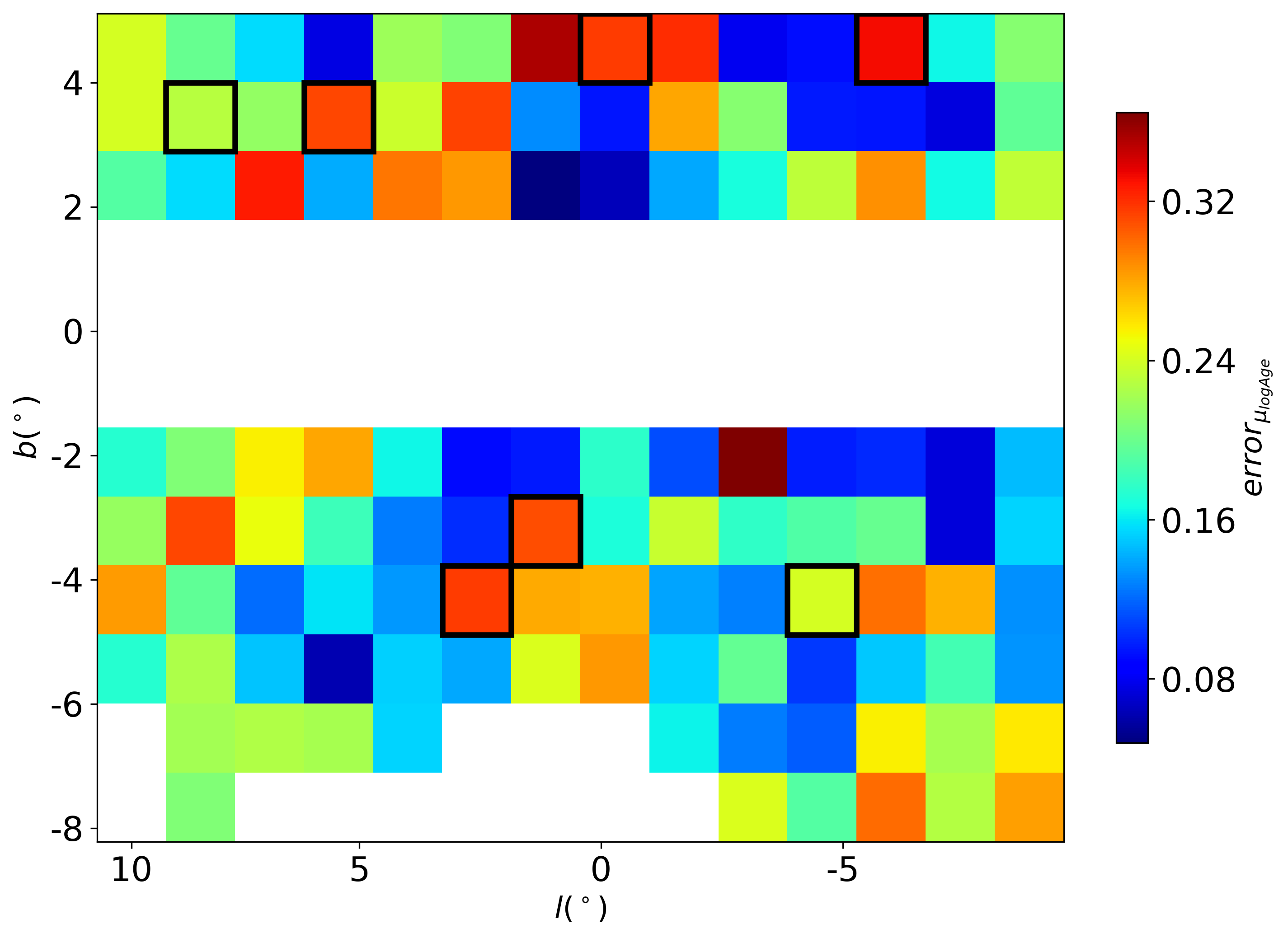}
\caption{}
\label{Fig.sub.4}
\end{subfigure}
\caption{(a) Map of the average logAge distribution for the Galactic Bulge. The blank regains are the tiles that are thrown away. It should be noted that due to extreme environmental effects, the MCMC results do not converge for regions where $|b| < 2^{\circ}$. (b) Error of the average values of the logAge. The non-convergence of MCMC results leads to significantly higher errors in the region $|b|<2^{\circ}$ where compared to other regions.
\label{fig:4.2-2}}
\end{figure}

\begin{figure}[ht!]
\centering
\includegraphics[scale=0.5]{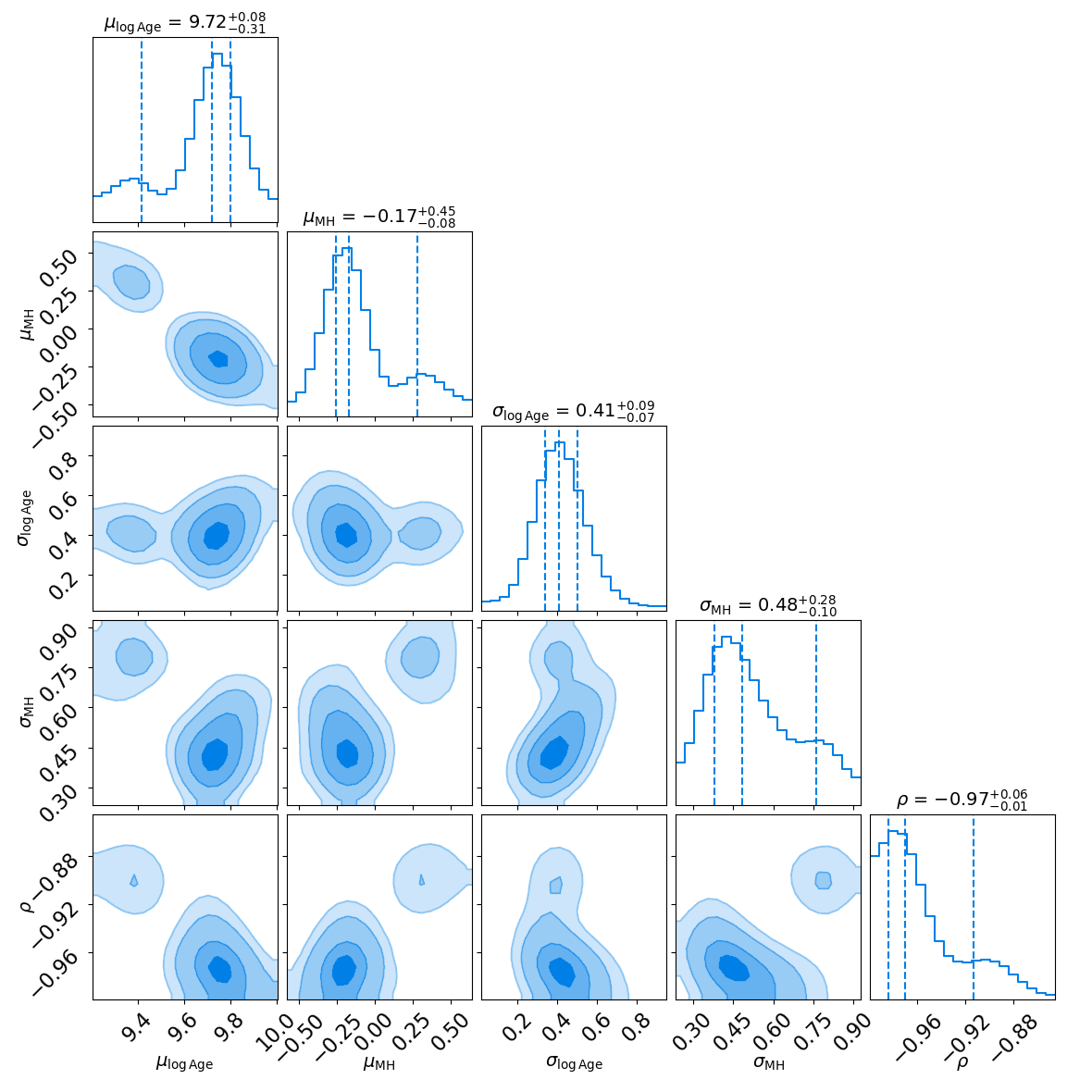}
\caption{Corner diagram of MCMC parameters for tile b389 ($l \sim-0.46^\circ$,$b \sim4.5^\circ$). Similar to Figure \ref{fig:4.1-1}
\label{fig:4.2-2-1}}
\end{figure}

Figure \ref{Fig.sub.3} shows the average logAge of the available tiles, and the location information is consistent with Figure \ref{fig:2.1}.
It can be seen from Figure \ref{Fig.sub.3} that the age of the $2^{\circ}<|b|<6^{\circ}$ region is significantly younger than that of the $|b|>6^{\circ}$ region. We explore these trends further in the following. Figure \ref{Fig.sub.4} shows the intrinsic error of the average logAge, whose calculation process is consistent with that of a single tile in Section \ref{sec:SF}. 
Visual inspection of the MCMC posterior plots revealed bimodal distributions in $\mu_{logAge}$ and $\mu_{[M/H]}$ for seven fields (b274, b279, b292, b379, b381, 385, b389), as shown in Figure \ref{fig:4.2-2-1}, which are marked with black squares in Figure \ref{fig:4.2-2}. These bimodalities introduce systematic offsets and larger intrinsic uncertainties in the average age. We attribute this to the observing quality of VISTA. Accordingly, when deriving the mean $\mu_{logAge}$ at each latitude, we used the median rather than the mean.

\begin{figure}[ht!]
\centering
\includegraphics[scale=0.25]{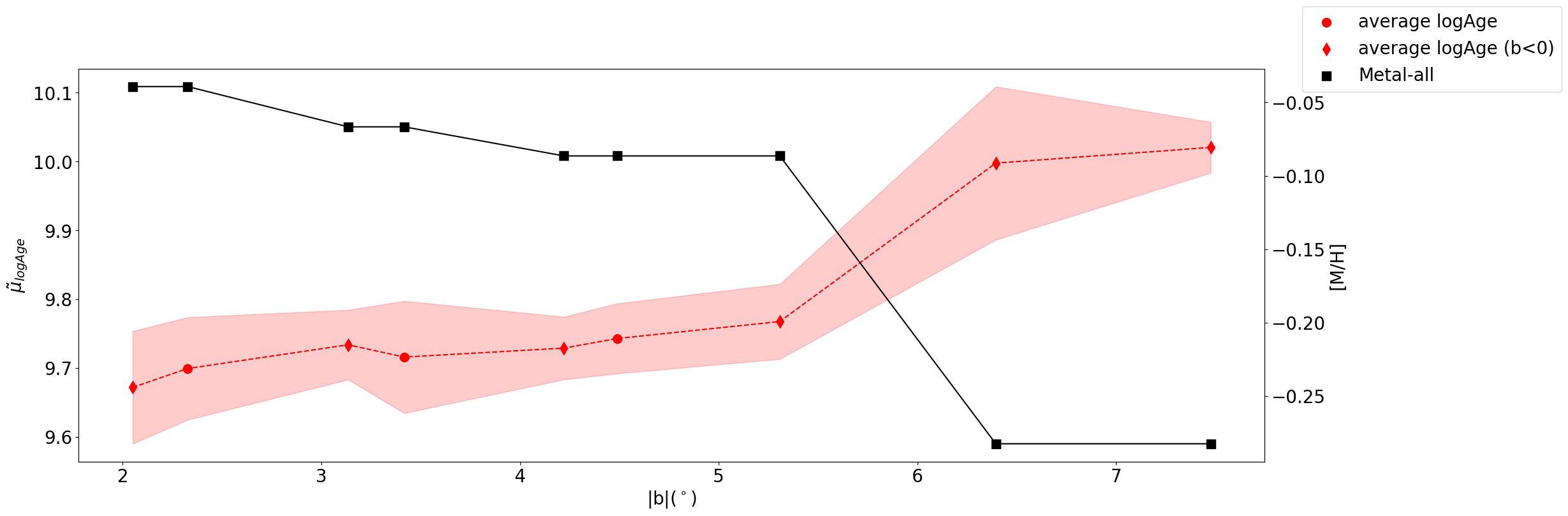}
\caption{The gradient of logAge and [M/H] with $b$. The left axis is logAge and [M/H] is on the right. The red dotted line represents the median value of the average logAge in the same $b$ and the red shadow is the dispersion of the average logAge in the same $b$. The circle shows the result for $b>$0 and the diamond shows the result for $b<$0. The straight line of black squares represents the prior on [M/H]. 
\label{fig:4.2-3}}
\end{figure}

To make it easier to see the relationship between age and $b$, we calculate the median (red dots in Figure \ref{fig:4.2-3}) and dispersion (red shading in Figure \ref{fig:4.2-3}) of the average logAge in the same $b$ and plot the relationship between logAge and the $|b|$.
The right axis of Figure \ref{fig:4.2-3} represents the values of [M/H], whereas the black straight line shows the relation between the [M/H] and $|b|$.

We find that logAge is monotonically increasing with Galactic latitude while [M/H] is monotonically decreasing. To check whether the variation of age depends on the variation of [M/H], we randomly select 9 test fields (b230, b248, b268, b280, b289, b312, b368, b372, b396) in the region $2^{\circ}<|b|<8^{\circ}$. We vary the prior distributions of [M/H] in these test fields, i.e., the distribution of [M/H] in $|b|>6^{\circ}$ is substituted for the distribution of $|b|<6^{\circ}$, while the distribution of $|b|<6^{\circ}$ is substituted for the distribution of $|b|>6^{\circ}$. 
We find that altering the metallicity prior introduces some fluctuations in the derived ages due to the age-metallicity degeneracy.
The prior [M/H] distribution varies by about 0.2 dex in mean. After swapping this prior, the average logAge shifts by roughly 0.1 dex, but the new results carry a larger intrinsic uncertainty of $\sim$0.15 dex. Taking this uncertainty into account, we conclude that the test field’s average logAge remains within the red shaded region of Figure \ref{fig:4.2-3}, even though the shaded span is approximately 0.05 dex.

As shown in the region $2^{\circ} < |b| < 8^{\circ}$, there is a clear positive gradient between logAge and $|b|$, which remains consistent regardless of the variations in [M/H]. This trend indicates a systematic transition in the average stellar age across the Galactic Bulge: the mean age increases from the near-plane regions (logAge $\sim 9.67 \pm 0.02$, $\text{age} \sim 4.69^{+0.97}_{-0.81}$ Gyr) toward the higher latitude regions (logAge $\sim 10.02 \pm 0.02$, $\text{age} \sim 10.48^{+0.93}_{-0.85}$ Gyr). The uncertainties for logAge are standard errors, calculated as $\sigma / \sqrt{N-1}$, where $N$ represents the number of tiles at each latitude. The corresponding errors in age are derived through error propagation from the logAge values. Consistently, for the lower latitude region ($2^\circ < |b| < 6^\circ$), the fraction of the age PDF younger than 5 Gyr is approximately 40\%$\sim$ 45\%. Whereas, for the higher latitude region ($|b| > 6^\circ$), this fraction drops to about 25\% $\sim$ 30\%.

However, it should be noted that the old ages found at higher latitudes ($|b| \sim 6^{\circ}$ to $8^{\circ}$, Fig. \ref{fig:4.2-3}) may be affected by the fact that we removed regions with double RC peaks. As explained in Section \ref{sec:AD}, these regions were left out because our current method for finding distances cannot determine a single average distance for such cases. Therefore, these areas could not be used in the isochrone fitting process. If the young stars are mostly part of the B/P structure, removing them might lead to an overestimation of the average age.

In terms of their physical origin, these two age groups likely represent different structural components. The younger population is closer to the Galactic centre, possibly contaminated by the star-forming region at the centre of the nuclear \citep{Lop01}, but most of these sources may have an origin from a pseudobulge originated from buckling instabilities of the disk 
\citep{Sel81,Nat17} or the long bar as a different component \citep{Lop07,Gon12,Amo13}. The oldest population is associated with a spheroidal/ellipsoidal bulge, usually called a ``classical bulge", formed through a monolithic collapse in the early times of the Galaxy formation \citep{Egg62,Obr13} or the accretion debris of the merged dwarf galaxies \citep{barnes_transformations_1992,obreja_two-phase_2012}. A clear look at their history requires a study of both chemistry and motion. We are now working on this detailed kinematic analysis and will share it in a future study.

Supporting this evolution, recent cosmological simulations of Milky Way-like galaxies such as Auriga and TNG50 provide a consistent ``in-situ" framework to explain such vertical age and metallicity structures. These models show that a process known as ``kinematic fractionation" \citep{fragkoudi_chemodynamics_2020} plays a key role in separating these populations. Under this mechanism, metal-poor stars have higher velocity dispersions that allow them to reach higher latitudes and form a more spread-out component. This theoretical distribution suggests that internal dynamics and evolution \citep[e.g.,][]{zhang_diverse_2025} are the main reasons for the bulge’s dual character \citep{grand_auriga_2017, gargiulo_high_2022}.

\subsection{Average [M/H] map of the Galactic Bulge}
\label{sec:GAM}

\begin{figure}[!htb]
\centering  %图片全局居中
\begin{subfigure}{0.5\textwidth}
\includegraphics[scale=0.3]{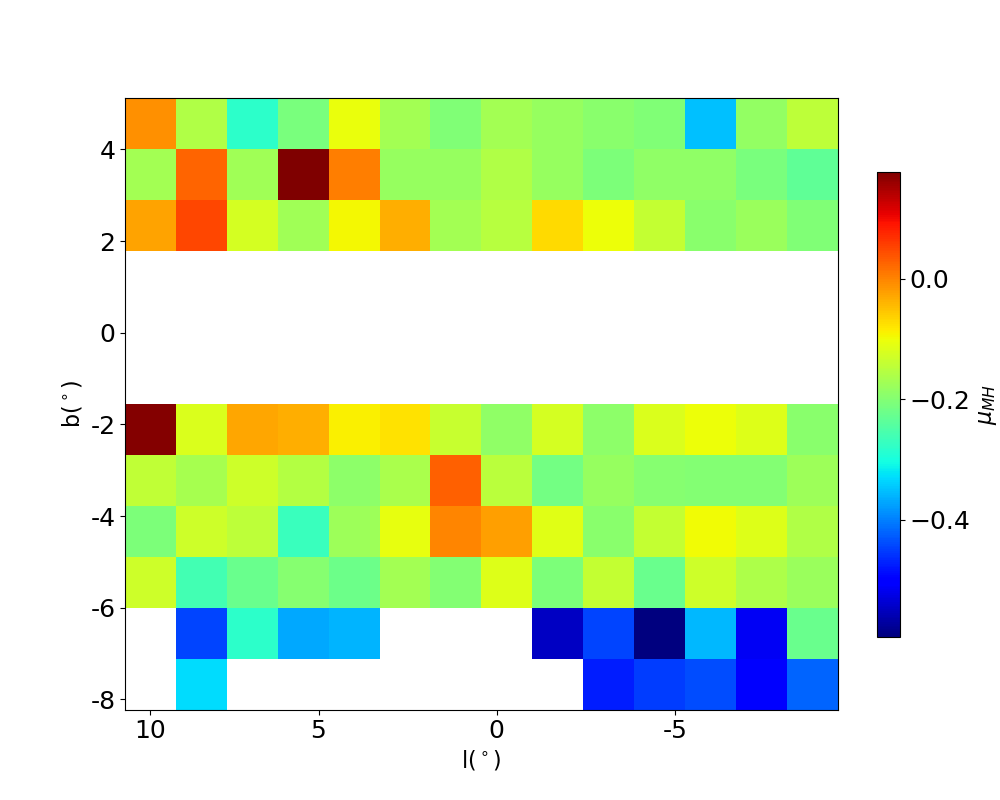}
\caption{}
\label{Fig.sub.4.3-1}
\end{subfigure}
\begin{subfigure}{0.48\textwidth}
\includegraphics[scale=0.3]{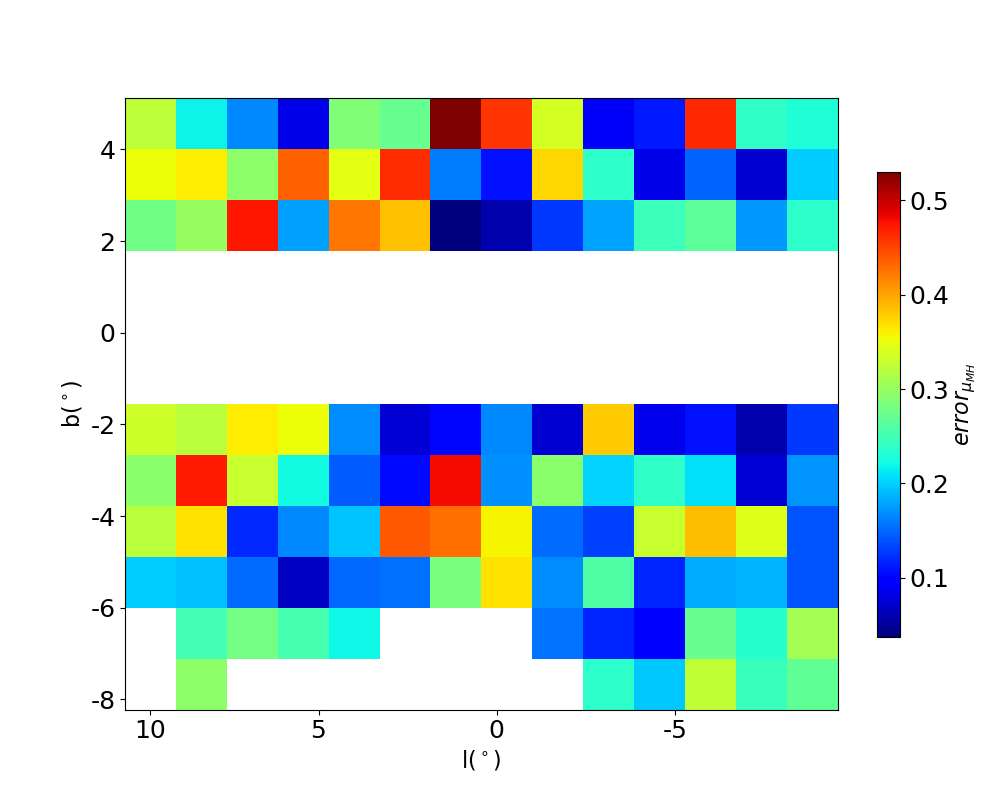}
\caption{}
\label{Fig.sub.4.3-2}
\end{subfigure}
\caption{Same as Fig. \ref{fig:4.2-2}, but for [M/H]. (a) Map of the average [M/H] distribution for the Galactic Bulge. (b) Error of the average values of [M/H].
\label{fig:4.3}}
\end{figure}

This section presents the global distribution of average [M/H] across the study area, derived using the methodology described in Section \ref{sec:IF}. As shown in Fig. \ref{fig:4.3}, the metal-poor populations are more concentrated in high-latitude regions ($|b| > 6^\circ$). It is important to note that these results are significantly influenced by the metallicity priors included in our model, which are based on the spectroscopic findings of \citet{rojas-arriagada_how_2020}. Our metallicity results remain sensitive to these model constraints.

\section{{Discussion}} 
\label{sec:Dis}
In this section, we assess the effects of distance uncertainties, foreground contamination, and sample selection bias. We also discuss the impact of stellar model uncertainties on our results and address the limitations of this study.

\subsection{Distance and distance uncertainty}
\label{sec:DisUn}
\begin{figure}[ht!]
\centering
\includegraphics[scale=0.3]{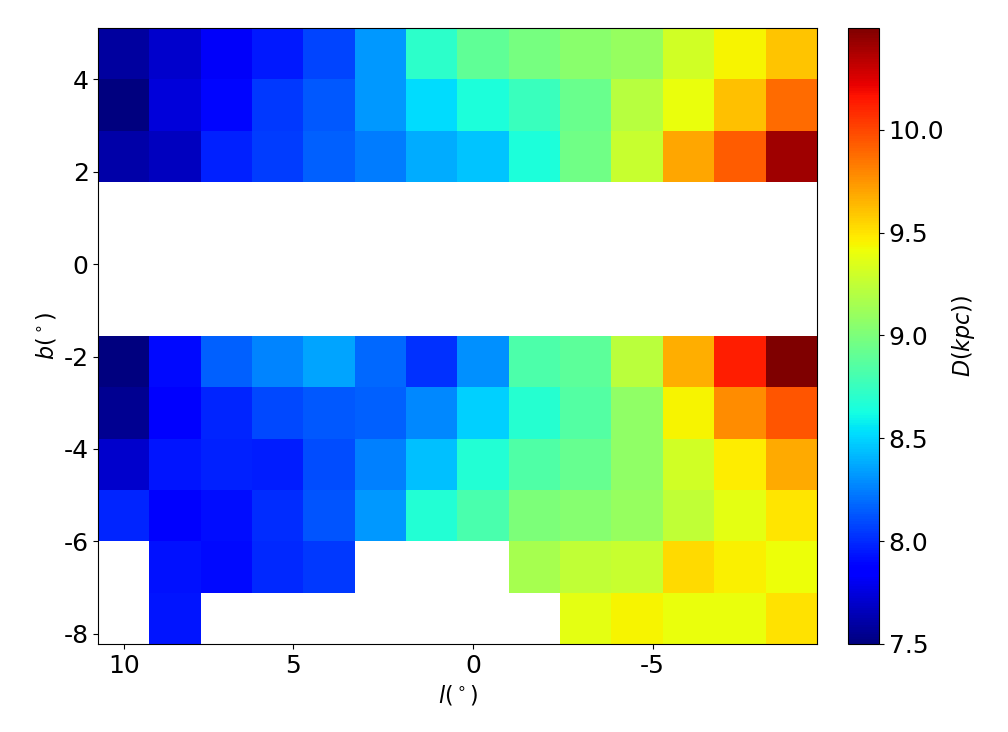}
\caption{Similar to Figure \ref{fig:4.2-2}, but showing the spatial distribution of the average distance. The blank areas also correspond to those in Figure \ref{fig:4.2-2}.
\label{fig:5.1-2}}
\end{figure}

Figure \ref{fig:5.1-2} shows the global average distances of the Galactic Bulge obtained using the method in Section \ref{sec:AD}. As can be seen from the figure, there is a gradient in distance along the $l$ direction, which is related to the depth of observation. However, the average logAge map (Figure \ref{fig:4.2-2}) does not show a gradient similar to that of distance, which indirectly demonstrates that the global average age obtained by this work are not affected by distance.

\begin{figure}[ht!]
\centering
\includegraphics[scale=0.2]{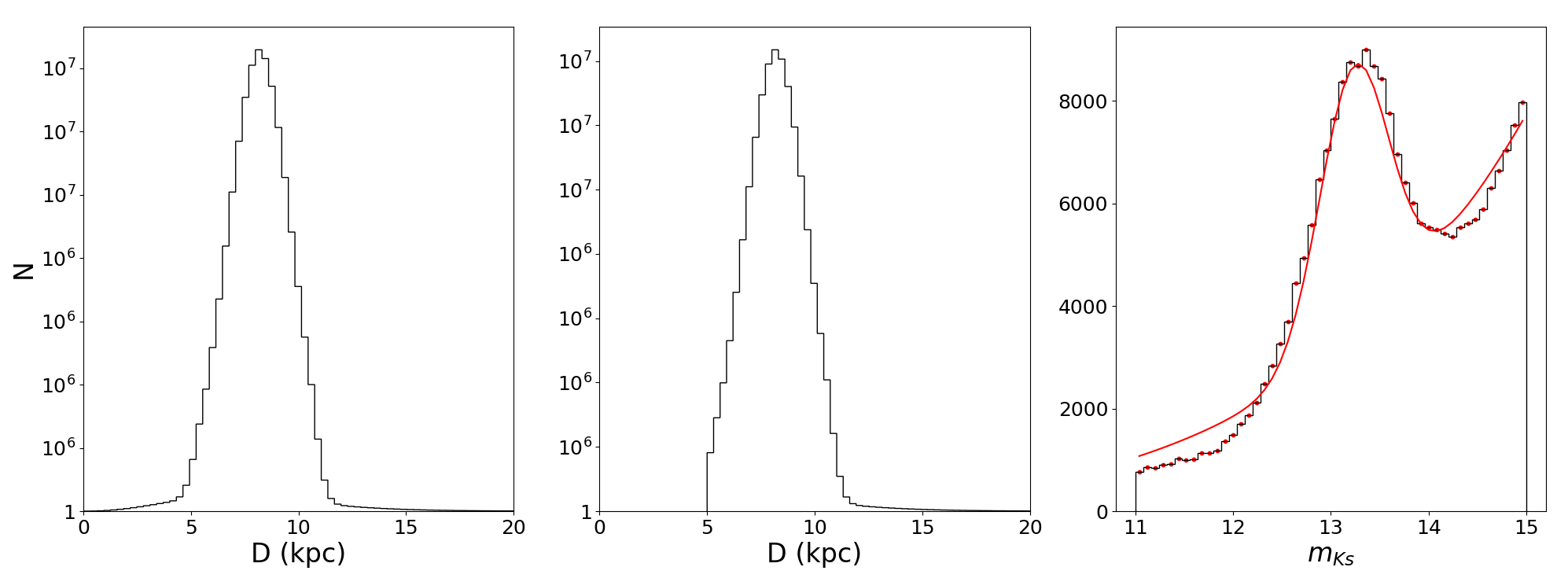}%Bulge_Test8.ebf
\caption{\textit{Left panel}: radial distance distribution for the full sample of GALAXIA simulations. \textit{Middle panel}: distribution of distances after throwing out stars within 5 kpc. \textit{Right panel}: $m_{K_{\mathrm{s}}}$ distribution of the RC stars, red dots are fitted points, the red line is the result of the fit.
\label{fig:5.1-1}}
\end{figure}

We use GALAXIA \citep{sharma_galaxia_2011} to generate a tile-sized sky region ($\sim$ 1.2 square degrees) centred on l = 0$^\circ$ and b = 4$^\circ$. It contains the foreground stars, Bulge member stars and background stars. The photometric system is DCMC ($JHK_{\mathrm{s}}$ form 2MASS). 
In performing the conversion from absolute to apparent magnitude, we add an error about the extinction, $\sigma_{A_{KS}}=0.482 \times \sigma_{E(J-K_{\mathrm{s}})}$.
$\sigma_{E(J-K_{\mathrm{s}})}$ is the extinction error from the MW-BULGE-PSFPHOT catalogues, and 0.482 from \cite{nishiyama_interstellar_2009}. To be close to the actual VVV observations, we throw out the stars fainter than the apparent magnitude $m_{K_{\mathrm{s}}}\sim$23. The distribution of line-of-sight distances for the final simulated sample is shown in the left panel of Figure \ref{fig:5.1-1}.

Similar to the average distance determination in the observed samples (more details in Section \ref{sec:AD}), we first throw out the stars within 5 kpc (the middle panel of Figure \ref{fig:5.1-1}), then, we fit the stellar magnitude distribution with $m_{ks}$ between 11 and 15 to get the mean $m_{K_{\mathrm{s}}}$ of the RC stars (the right panel of Figure \ref{fig:5.1-1}). Finally, we obtain the average distance according to Eq. \ref{eqn:2}. The absolute magnitude of the RC star in GALAXIA is based on the \cite{sharma_galaxia_2011}, which is $M_{K_{\mathrm{s}}}=$-1.5. 

We found that the average distance for the simulated samples is overestimated. The difference between the average distance and the ground truth is about 8\%. To evaluate the errors in our average distance calculation process, we use the mean radial distance of the stars in the middle panel of Figure \ref{fig:5.1-1} as the true average distance for the Bulge (the ground truth). These values are from the catalog provided by the GALAXIA simulation.
This suggests that the systematic uncertainty in calculating the distances is approximately 8\% of the actual distance.
This uncertainty is consistent with an expected methodological bias: the peak of the observed magnitude distribution, N(m), which we use to determine distances, does not perfectly coincide with the peak of the underlying stellar density distribution, $\rho(d)$, due to the volume effect ($N(m(d)) \propto \rho(d) \cdot d^{3}$). This small shift, which scales with the intrinsic dispersion of the red clump absolute magnitude ($\sigma_{RC}$), explains the observed discrepancy. As this distance bias corresponds to a negligible shift of only $\sim$0.16 mag (according to Eq. (\ref{eqn:2})), it does not significantly impact our subsequent age determinations.

Following an estimated 8\% distance uncertainty from the GALAXIA sample, we apply the Section \ref{sec:IF} method to simulated data to assess its effect on our age results.
First, we convert the absolute magnitudes of the GALAXIA simulated stars into apparent magnitudes by using each star's true distance and adding both extinction and photometric errors. Similar to the method in Section \ref{sec:IF}, we select a subset of stars with $J-K_s \in [0.4, 0.7]$ and $M_{K_s} \in [0.8, 1.4]$ (consistent with the black box in Fig. \ref{fig:3.1}) as the sample for average age fitting. In this step, the distance modulus is set to the average distance previously estimated from RC stars. All other process follows Section \ref{sec:IF}, except that we replace the truncated Gaussian function in Equation (\ref{eqn:4-6}) with a Gaussian Mixture Model (GMM) consisting of three components. This change is necessary because the intrinsic color ($J-K_s$) distribution in GALAXIA (without observational errors) shows a non-Gaussian profile related to its simulation process. To investigate the impact of the average distance on our final results, we convert the apparent magnitudes of the sample into absolute magnitudes using the calculated distance modulus, rather than the true values before fitting the isochrones.

\begin{figure}[ht!]
\centering
\includegraphics[scale=0.5]{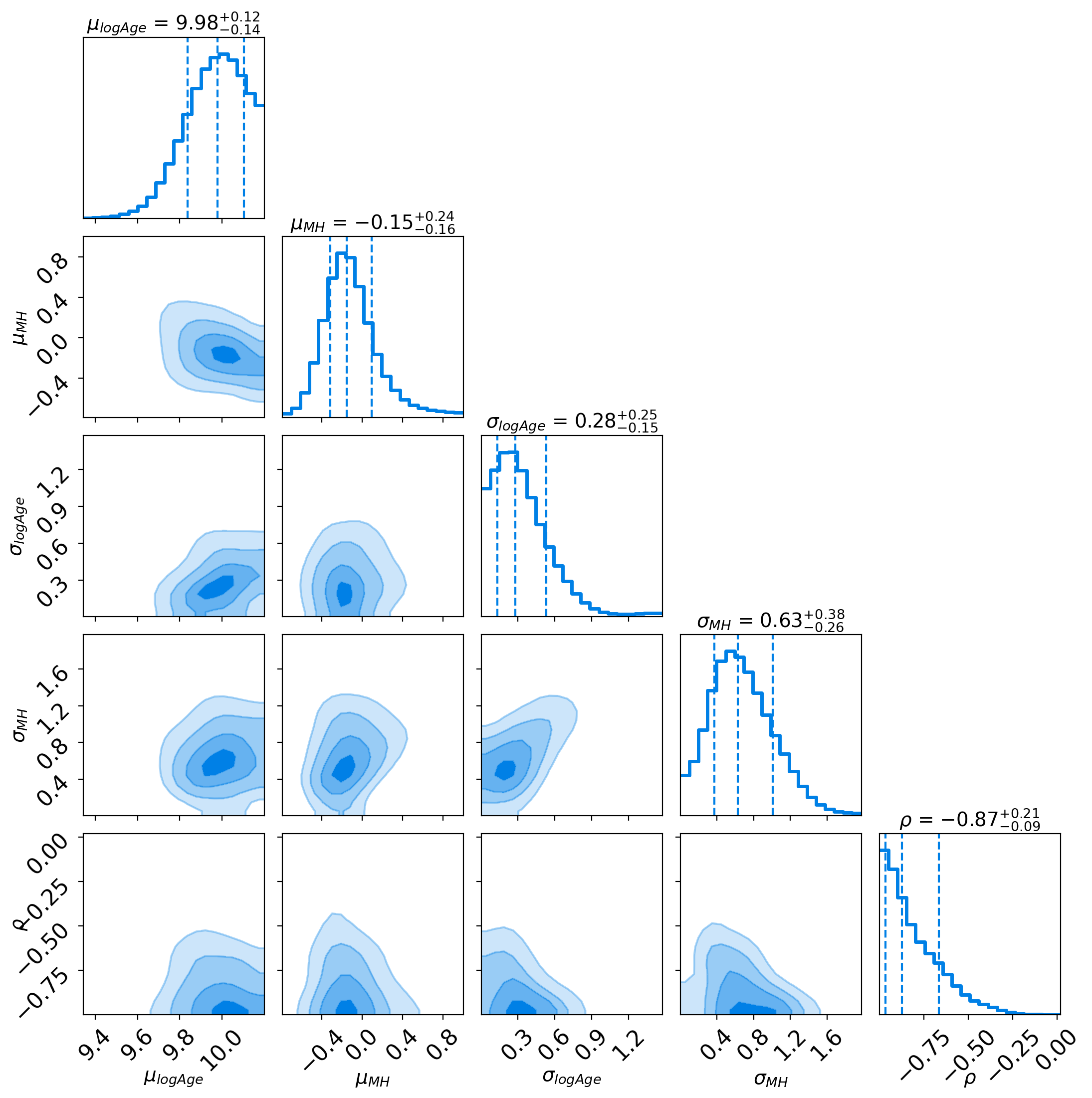}%Bulge_Test8.ebf
\caption{Same as Figure \ref{fig:4.1-1}, but showing the results for the GALAXIA simulation data.
\label{fig:5.1-3}}
\end{figure}

Figure \ref{fig:5.1-3} displays the final fitting results, where the mean logAge is 9.98 and [M/H] is -0.15. For comparison, the true mean and median logAge of the GALAXIA sample are 9.99 and 10.0, respectively, while the true mean and median [M/H] are -0.15 and -0.12. It is important to note that the [M/H] prior used in this test is a Gaussian distribution fitted to the true values of the simulated data. The high consistency between our recovered parameters and the ground truth demonstrates that the line-of-sight depth and the assumption of a uniform average distance do not induce significant biases into our average age determinations. 

For consistency, we also perform a test on the observed data using tile b393, replacing the truncated Gaussian function in Equation (\ref{eqn:4-6}) with the similar GMM used in the GALAXIA test. We find that while the MCMC parameter results remain similar to those in Figure \ref{fig:4.1-1}, the intrinsic uncertainties of the parameters are significantly larger. This is because a single truncated Gaussian function is already sufficient to describe the color distribution of the observed data; introducing a GMM instead leads to overfitting.

In light of the test based on simulated data, we assess the robustness of the average age estimation using the observed data. We change the distances by $\pm$10\% for an individual tile and find that the new logAge estimates do not exceed 1$\sigma$ uncertainty. This indicates that our results of the average logAge do not significantly change when the distance is systematically changed by 10\%.

\subsection{Foreground contamination}
In this work, we use Gaia DR3 parallaxes to exclude foreground stars (details in Section 3.1). To evaluate the residual contamination from remaining foreground disk stars in the sample after applying this filtering method, we utilize the same GALAXIA simulation sky area used in the previous section.

Prior to evaluation, we incorporate Gaia distance measurement errors into the stellar distances from the GALAXIA simulation. These errors are modeled as a Gaussian distribution derived by converting the mean and variance of parallax errors of foreground stars which defined in Eq. (\ref{eqn:1}) of Section \ref{sec:SS} from the cross-matched photometric catalog with Gaia DR3 into distance errors. However, no constraints are applied to the parallax errors themselves; that is, the final sub-equation within Eq. (\ref{eqn:1}) is not considered.

The GALAXIA simulation provides the stellar population composition for each star. After applying our foreground star selection criteria to remove foreground objects, the resulting sample contains approximately 5\% thin disk stars younger than 10 Gyr, while bulge member stars constitute about 90\% of the population.

To further verify the robustness of our results against the key finding of young stars, we perform a sensitivity test on tile b393 by adopting a more aggressive parallax cut, $\varpi - \varpi_0 > 0.1$ mas. The MCMC results for the stellar parameters remain similar to our original findings (consistent with Fig. \ref{fig:4.1-1}), with the mean values of logAge and [M/H] showing no significant shifts. This suggests that our conclusions regarding the young stellar population at lower latitudes are not significantly biased by foreground contamination.

\subsection{Completeness and selection effects}
\label{sec:Com&SE}
To account for sample selection effects, we have incorporated the completeness of the MW-BULGE-PSFPHOT catalog directly into our modeling. 
This completeness was determined by \citet{surot_mapping_2019_2} through artificial star tests. In our method, we apply the average completeness values from the catalog as weights within the likelihood function for each magnitude bin (see Section \ref{sec:IF} for details). This approach ensures that the model accounts for the relative loss of stars when fitting the color distribution.

\begin{figure}[ht!]
\centering
\includegraphics[scale=0.5]{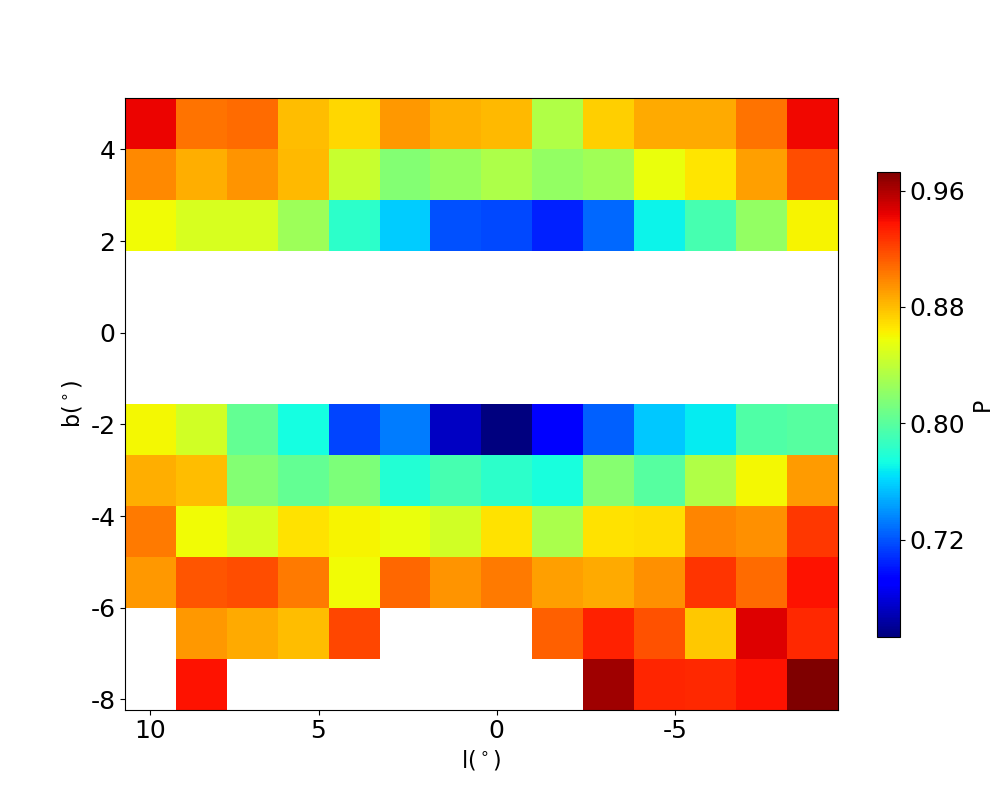}
\caption{Global distribution of the mean completeness for the stellar samples used in age fitting for each tile. The completeness values are derived from artificial star tests by \citet{surot_mapping_2019_2} and are available in the published MW-BULGE-PSFPHOT catalog.
\label{fig:5.3}}
\end{figure}

As shown in Fig. \ref{fig:5.3}, the global distribution of the average completeness for our RG sample exceeds 80\% in the majority of the studied regions. The combination of high sample completeness and the inclusion of completeness in our model ensures that the results are not significantly biased by selection effects.

\subsection{Stellar model uncertainty}
Isochrone models are constructed based on a theoretical understanding of stellar evolution. However, these theories are not perfect.
\citet{How19}  revealed significant limitations in the use of theoretical models of stellar isochrones to derive age and metallicity. They claim that the uncertainty of the stellar models should be quantitatively taken into account. These include convection in stellar models, stellar opacities and atomic diffusion during main sequence star evolution. Stellar opacities are something to consider for stars whose tracers are below the main-sequence star. Atomic diffusion occurs during the evolution of the main-sequence star.

Convection in stellar models is about stellar atmosphere modeling. \cite{How19} found that the $\alpha_{MLT}$ in the models (e.g. PARSEC) do not describe all stars; in particular, one of the values of $\alpha_{MLT}$ chosen for the Sun does not apply to stars on the giant branch. But $\alpha_{MLT}$-related research has been inconclusive in terms of accurate results. The current combination of asteroseismology and spectroscopy has yielded better-fitting results \citep{tayar_correlation_2017}.

In addition to this, \cite{nataf_reconciling_2012} showed that the chemically evolved helium abundance in the Galactic Bulge is higher compared to the standard isochron model. They concluded that the low helium abundance used in the standard isochrones leads to an uncertainty of about 30\% in the inferred stellar age.
These may affect the intrinsic error of our results. 
We hope to be able to use more accurate models in the future.

\subsection{Limitations}
Our study combines infrared photometry with spectroscopic data to determine the average age of the Galactic bulge. As shown by the likelihood function (Eq. \ref{eqn:9}), we aim to reconcile the stellar colors from photometry and the metallicities from spectroscopy with a two–dimensional Gaussian defined by logAge and [M/H]. However, as noted by \cite{queiroz_bulge_2020} and \cite{queiroz_milky_2021}, the ages derived from the spectroscopic dataset currently differ markedly from those inferred from the photometry. This discrepancy is partly due to selection effects. Because of the heavy extinction toward the bulge, the spectra preferentially contain brighter, younger stars, biasing the metallicity prior toward higher values and driving our final results toward younger results.

Another limitation of the model is that we simplify the logAge–[M/H] relation as a simple two-dimensional Gaussian. At the evolutionary stage of RG stars, age and metallicity are highly degenerate, and this degeneracy does affect our results. However, given the photometric errors and color accuracy of the VVV data, we do not reach the main-sequence turn-off stars depth. We anticipate that the forthcoming, wide-field, high-resolution Chinese Space Station Survey Telescope (CSST) will provide deeper insights into the Galactic bulge.

\section{SUMMARY} \label{sec:Sum}

We used the highly accurate PSF-fitting photometric catalogs for $J$ and $K_{\mathrm{s}}$ bands of VISTA in the Galactic Bulge to map the distribution of the average age of stars across the Galactic Bulge.
The RG stars were used as tracers, and their average distances were calculated using the RC stars as standards; their average ages were then determined using PARSEC's isochrones. 

We summarised as: we found a significant systematic age gradient across the Galactic Bulge ($2^{\circ} < |b| < 8^{\circ}$). The average stellar age exhibits a clear latitudinal dependence, increasing from $\sim 4.69^{+0.97}_{-0.81}$ Gyr at lower latitudes to $\sim 10.48^{+0.93}_{-0.85}$ Gyr at higher latitudes. 
This trend indicates there are more young stars at low latitudes and more old stars at high latitudes.

We suggest that this may be related to the chemical-kinetic evolution of the Bulge. The predominance of young stars at low latitudes may be an effect of star-forming regions and most of them may have an origin from a pseudobulge originated from buckling instabilities of the disk or the long bar as a different component. The predominance of old stars at high latitudes may be related to their dynamics. This is associated with a spheroidal/ellipsoidal bulge, ``the classical bulge", formed through a monolithic collapse in the early times of the Galaxy formation or the accretion debris of the merged dwarf galaxies. We also reference cosmological simulations of Milky Way-like galaxies to provide further evidence for our perspective.

Finally, we evaluated the impact of distance uncertainties, foreground contamination, and sample selection bias. We also reviewed the role of stellar model uncertainties and addressed the current limitations of this work.

\begin{acknowledgements}
This work is supported by the National Key R\&D Program of China No. 2025YFF0511000.
M.L.C. is supported by the Chinese Academy of Sciences President's International Fellowship Initiative grant No. 2023VMB0001 and grant PID2021-129031NB-I00 of the Spanish Ministry of Science (MICIN). H.F.W is supported in this work by the Department of Physics and Astronomy of Padova University through the 2022 ARPE grant: {\it Rediscovering our Galaxy with machines.} The authors acknowledge with gratitude the use of data from ESO's Public Survey program \texttt{179.B-2002}, collected using the VISTA telescope.
\end{acknowledgements}

\bibliographystyle{raa}
\bibliography{Reference_list}
\label{lastpage}

% \end{CJK}
\end{document}